\documentclass[useAMS,usenatbib,referee]{mn2e}
\usepackage{multirow}
\usepackage[dvipdfmx]{graphicx}
\usepackage{amsmath}	
\usepackage{amssymb}	
\usepackage{float}

\newcommand{\msun}{\thinspace M_\odot}

\newcommand{\vect}[1]{\mbox{\boldmath$#1$}}
\def\lesssim{\mathrel{\hbox{\rlap{\hbox{\lower4pt\hbox{$\sim$}}}\hbox{$<$}}}}
\def\gtrsim{\mathrel{\hbox{\rlap{\hbox{\lower4pt\hbox{$\sim$}}}\hbox{$>$}}}}
\newcommand{\cm}{\,{\rm cm}^{-3} } 
\newcommand{\gcm}{\,{\rm g}\, {\rm cm}^{-3} } 
\newcommand{\km}{\,{\rm km\, s}^{-1}}

\newcommand{\dg}{^\circ}

\title[Misalignment of Magnetic Fields]{Misalignment of Magnetic Fields, Outflows and Discs in Star-forming Clouds}
\author[Machida et al.]{
Masahiro N. Machida,$^{1}$\thanks{E-mail: machida.masahiro.018@m.kyushu-u.ac.jp (MNM)} 
Shingo Hirano$^{1}$
and 
Hideyuki Kitta$^{1}$
\\
Department of Earth and Planetary Sciences, Faculty of Sciences, Kyushu University, Fukuoka, Fukuoka 819-0395, Japan
}
\date{ }
\pubyear{2019}
\begin{document}
\label{firstpage}
\pagerange{\pageref{firstpage}--\pageref{lastpage}}
\maketitle

\begin{abstract}
Using resistive magnetohydrodynamics simulations, the propagation of protostellar jets, the formation of circumstellar discs and the configuration of magnetic fields are investigated from the prestellar cloud phase  until $\sim$500\,yr after protostar formation. 
As the initial state, we prepare magnetized rotating clouds, in which the rotation axis is misaligned with the global magnetic field by an angle  $\theta_0$.
We calculate the cloud evolution for nine models with different $\theta_0 (=$ 0, 5, 10, 30, 45, 60, 80, 85, 90$\dg$).
Our simulations  show that there is no significant difference in the physical quantities of the protostellar jet, such as the mass and momentum, among the models except for the model with $\theta_0=90\dg$. 
On the other hand, the directions of the jet, disc normal and magnetic field are never aligned with each other during the early phase of star formation except for the model with $\theta_0=0\dg$.
Even when the rotation axis of the prestellar cloud is slightly inclined to the global magnetic field, the directions of the jet, disc  normal and local magnetic field differ considerably, and they randomly change over time.
Our results indicate that it is very difficult to extract any information from the observations of the directions of the outflow, disc   and magnetic field at the scale of $\lesssim 1000$\,au.
Thus,  we cannot use such observations to derive any restrictions on the star formation process.
\end{abstract}

\section{Introduction}
\label{sec:intro}
Magnetic fields and rotation play significant roles in the star formation process. 
The rotation of a prestellar cloud results in the formation of a Keplerian disc,  which is the host of planet formation \citep{hayashi85}.
The presence of a magnetic field leads to the emergence of  protostellar jets, which transfer an excess angular momentum and determine the star formation efficiency and the final stellar mass \citep{nakano85}.
Since magnetic fields and angular momentum are vectors, their directions should be considered.
For example, the efficiency of  magnetic braking depends on the mutual angle between the magnetic and angular momentum vectors of the prestellar cloud {\citep{mouschovias79,mouschovias80}.
In addition, the directions of the magnetic and angular momentum vectors influence the disc  formation process \citep{matsumoto04,hennebelle09,tsukamoto18}. 

In many theoretical studies,  the magnetic fields are assumed to be parallel to the rotation axis for simplicity \citep[][and references therein]{larson03}. 
In numerical simulations, researchers have calculated the evolution of rotating magnetized clouds assuming that the rotation axis is parallel to the global magnetic field \citep{tomisaka02,machida04,banerjee06,tomida13,bate14,vaytet18}.

\citet{matsumoto04} first investigated the misalignment case, in which the rotation axis of the prestellar cloud is inclined to the global magnetic field.
They showed that the directions of the local magnetic field and disc  normal (or local angular vector) tended to converge.
Although they did not show perfect alignment, the outflow direction roughly agrees with the direction of the local magnetic field, not the direction of the global magnetic field.

\citet{hennebelle09} also investigated cloud collapse when the global magnetic field is not parallel to the cloud rotation axis. 
Although they did not comment on the outflow direction, the outflow seems not to be aligned with the initial magnetic field. 
In these studies, the calculations were executed in the ideal  magnetohydrodynamics (MHD) approximation.
However, in reality, dissipation of the magnetic field can significantly reduce the efficiency of magnetic braking  \citep{machida11}. 
Since the direction of the rotationally supported disc  (or the disc  normal direction) is changed by magnetic braking  \citep{joos12},  and the disc  drives protostellar jets,  the magnetic dissipation should be considered when investigating the directions of the disc  and jets. 

Using non-ideal MHD simulations, \citet{masson16} and \citet{tsukamoto18} investigated cloud evolution for a rotation axis that was not aligned with the global magnetic field (i.e. the misalignment case) and discussed the size growth of the circumstellar disc.
They calculated the gas collapsing phase just before and after protostar formation.
Thus, we cannot know the directions of the evolved protostellar jet and disc  from their simulations.
\citet{wurster19} investigated disc  formation and fragmentation with the sink particle technique in non-ideal MHD simulations. 
They were able to calculate the long-term evolution during the main accretion phase, while the high-speed jet does not emerge in their simulations  due to adopting a large sink radius of 1\,au. 
 
\citet{hull13} noted that observed magnetic fields at a large scale are not aligned with the outflow directions.
The speed of observed outflows exceeds $ 10\km$ with a scale of $\sim 1000$\,au.
In contrast, in many simulations, the outflow speed does not reach $\sim10\km$ \citep{matsumoto04,machida05b,hennebelle09,seifried12,wurster19}.
Thus, the outflows in past simulations are much slower than those in  observations. 
Only low-velocity outflows appear in the simulations either because the spatial resolution of the simulations is very low or because the time integration is not sufficient to evolve high-speed flows.
Whatever the reason, there is currently a significant gap between past simulations and observations with regard to the speed of outflows.

Recent developments in super computers enable the calculation of long term cloud evolution with a high spatial resolution.  
Recent studies have calculated an outflow reaching $\sim1000$\,au with a speed of $\gtrsim 100\km$, in which a protostar can be resolved with a sufficiently high spatial resolution \citep{machida14,machida19}.
In addition, \citet[][hereafter Paper I]{hirano19} showed that the outflow directions differ for different speeds, so that the low-velocity component is misaligned with the high-velocity component, which was confirmed in a recent observation \citep{matsushita19}.
We can now reasonably compare simulations  with observations.

The purpose of this study is to verify the star formation process in numerical simulations compared with observations.
We focus on the directions of the outflow, disc  and magnetic field  in the early phase of star formation.
Our paper is structured as follows. 
We present the initial conditions and numerical settings in \S2.
The simulation results are shown in \S3. 
We quantitatively discuss differences between the alignment and misalignment cases in \S4. We summarize our results in \S5.

\section{Initial Conditions and Numerical Settings}
\label{sec:settings}
The initial conditions and numerical settings in this study are the same as those in Paper I, so we only briefly describe them here.
As the initial state, we adopt a Bonnor--Ebert sphere with  a central density of $6\times10^5\cm$ (or $2.4\times10^{-18}\gcm$) and an isothermal temperature of $10$\,K. 
To promote cloud contraction, the density of the Bonnor--Ebert sphere is uniformly enhanced by a factor of 1.8 \citep{machida13}.
The initial cloud has a radius of $1.18\times10^4$\,au and a mass of $2.1\msun$.
A rigid rotation of $\Omega_0=1.4\times10^{-13}$\,s$^{-1}$ is set inside the initial cloud, while a uniform magnetic field of $B_0=5.7\times10^{-5}$\,G is imposed over the whole computational domain.
The ratios of the thermal ($\alpha_0$) and rotational ($\beta_0$) energies to the gravitational energy are  $\alpha_0=0.39$  and $\beta_0=0.026$.
The mass-to-flux ratio normalized by the critical value $(2\pi G^{1/2})^{-1}$ is $\mu_0=1.2$.
These non-dimensional parameters  ($\alpha_0$, $\beta_0$ and $\mu_0$) are the same as those adopted in \citet{machida14}, \citet{machida19} and Paper I.

We define the initial mutual angle $\theta_0$ between the rotation axis of the initial cloud and the global magnetic field, and use it as a parameter.
Each model is characterized only by the parameter $\theta_0$.
We take nine different initial angles $\theta_0=0$, 5, 10, 30, 45, 60, 80, 85 and 90$\dg$ and execute calculations for these nine models.
The model names and parameters are summarized in Table~\ref{table:1}.

To calculate the cloud evolution for these models, we use our nested grid code, in which rectangular grids  are nested \citep[for details, see][]{machida04,machida05,machida07,machida11,machida14}. 
Each grid is composed of ($i$, $j$, $k$) = (64, 64, 64) cells.
In Cartesian coordinates, the uniform magnetic fields are set along the $z$-axis, while the rotation axis is inclined from the $z$-axis to the $x$-axis by the angle $\theta_0$. 

Each grid level is described by $l$, and the minimum  and maximum  grid levels are set to $l=1$, which corresponds to the coarsest grid, and $l=20$, which  corresponds to  the finest  grid, respectively. 
The grid size and cell width halve with each increment of grid level.
The grid size $L (l)$ and cell width $h(l)$ of the coarsest grid ($l=1$) are $L(1)=3.78\times10^5$\,au and $h(1)=591$\,au,
while the finest or maximum grid ($l=20$) has   $L(20)=0.72$\,au and $h(20)=0.011$\,au. 

The initial cloud is immersed in the fourth level grid ($l=4$), whose grid size is equal to twice the Bonner--Ebert radius ($2.36\times10^4$\,au).
A uniform density of $n_{\rm ISM}=1.35\times10^4\cm$ is set for the exterior of the initial cloud, and the numerical boundary is set to the surface of the $l=1$ grid. 
Thus, outside the initial cloud, the interstellar medium is distributed in the range of $1.18\times10^4\,{\rm au} < r < 1.89\times10^5\, {\rm au}$.
A large area of interstellar space is used to prevent any artificial reflection of the Alfv\'en wave at the boundary \citep{machida13}.
A finer grid is automatically generated to resolve the Jeans wavelength, of at least 16 cells.

The basic equations used in this study are as follows:
\begin{eqnarray} 
& \dfrac{\partial \rho}{\partial t}  + \nabla \cdot (\rho \vect{v}) = 0, & \\
& \rho \dfrac{\partial \vect{v}}{\partial t} 
    + \rho(\vect{v} \cdot \nabla)\vect{v} =
    - \nabla P - \dfrac{1}{4 \pi} \vect{B} \times (\nabla \times \vect{B})
    - \rho \nabla \phi, & \\ 
& \dfrac{\partial \vect{B}}{\partial t} = 
   \nabla \times \left[  \vect{v}  \times \vect{B} - \eta_{O} (\vect{\nabla} \times \vect{B})   \right], & \\
& \nabla^2 \phi = 4 \pi G \rho, &
\end{eqnarray}
where $\rho$, $\vect{v}$, $P$, $\vect{B} $, $\eta_{O}$ and $\phi$ denote the density, velocity, pressure, magnetic flux density, Ohmic resistivity and gravitational potential, respectively.
We use the fitting fomura of the Ohmic resistivity $\eta_O$, which was derived in \citet{machida07}  and is described as  \begin{equation}
\eta_{O} = \dfrac{740}{X_e}\sqrt{\dfrac{T}{10 {\rm K}}} \left[ 1-{\rm tanh}\left( \dfrac{n}{10^{15}\cm}  \right)  \right] \, \, \, {\rm cm}^2\,{\rm s}^{-1},
\label{eq:etadef}
\end{equation}
where $T$ and $n$ are the gas temperature and the number density of molecular hydrogen, and $X_e$ is the ionization degree of the gas which describes as
\begin{equation}
X_e =  5.7 \times 10^{-4} \left( \dfrac{n}{{\rm cm}^{-3}} \right)^{-1}.
\label{eq:xe}
\end{equation}
Further  information about equations~(\ref{eq:etadef})  and (\ref{eq:xe}) can be referred to \citet{nakano02}. 
For the gas pressure and temperature,  the barotropic equation of state $P\propto \rho^\Gamma$ (or $T\propto \rho^{\Gamma-1}$) is used \citep[for detals, see][]{machida07,machida18}.
The polytropic exponent $\Gamma$ is set as
\begin{equation} 
\Gamma = \left\{
\begin{array}{lll}
 1    & {\rm for} &   \rho < 8.0 \times 10^{-14} \gcm\\
 5/3  & {\rm for} &  8.0 \times 10^{-14} \gcm < \rho < 1.2\times10^{-10} \gcm , \\
 7/5  & {\rm for} & 1.2 \times 10^{-10} \gcm < \rho < 4.0\times10^{-9} \gcm , \\
 1.1  & {\rm for} &  4.0 \times 10^{-9} \gcm < \rho < 1.2\times10^{-5} \gcm \\
 2    &  {\rm for} & \rho > 1.2\times10^{-5} \gcm.
\label{eq:gamma}
\end{array}
\right.  
\end{equation}
The equation of state in the range $\rho < 1.2\times10^{-5} \gcm $ approximately reproduces the thermal evolution shown in past studies \citep{larson69,masunaga00,tomida13}.
On the other hand, in order to realize a long-term integration, a stiff equation of state (or larger $\Gamma$) is adopted in the range of $\rho > 1.2\times10^{-5} \gcm$, which also mimics the protostar having  a size of $\ll 10$ solar radius  \citep[for details, see][]{machida14,machida15,machida19,hirano19}. 
Thus, the gas temperature is only a function of density with the barotropic approximation, and  the change in temerature due to such as shock heating and protostellar irradiation is ignored.
Since we only focus on the very early phase of star formation (see \S\ref{sec:results}), the heating (and cooling) effects are expected to not be significant.
However, we need to carefully consider the gas temperature because it  is related to the ionization degree and Ohmic resistivity which detemine the configuration of magnetic field in a small scale. 
We willl focus on the effect of heating and cooling on the magnetic configuration in future studies. 
The numerical methods and settings are fully described in our previous studies \citep{machida04, machida05, machida06,machida13, machida14, machida18, machida19, hirano19}.

\section{Results}
\label{sec:results}
We calculated the cloud evolution from the prestellar cloud stage until about 500\,yr after protostar formation for the nine models. 
This section describes the cloud evolutions for the models T00 and T90 and then presents the results for the models T05 to T85. 

\subsection{Rotation Axis Parallel to Magnetic Field: $\theta_0=0\dg$}
Figure~\ref{fig:1} shows the density and velocity distributions on the $x=0$, $y=0$ and $z=0$ cutting planes at $t_{\rm ps}=462.2$\,yr for model T00, where $t_{\rm ps}$ is the elapsed time after protostar formation.
The protostar has a mass of $\sim0.033\msun$ at this epoch.
\footnote{
We define the protostar as the region where the number density exceeds  $n>10^18\cm$ as described in Paper I. 
}
For this model, both the initial magnetic field  and initial rotation vectors are set to be parallel to the $z-$axis.
As typically seen in past simulations \citep{bate98,machida14,tomida15}, a rotationally supported disc  forms around the protostar (Fig.~\ref{fig:1}{\it i}).
The disc  has a size of $\sim5$\,au (Figs.~\ref{fig:1}{\it d} and {\it e}). 
As also seen in past studies \citep{tomisaka02, banerjee06, machida06}, both low- and high-velocity flows appear above and below the disc. 
The low-velocity flow has a wide opening angle, while the high-velocity component has a narrow structure (Figs.~\ref{fig:1}{\it a}, {\it b}, {\it d} and {\it e}). 
We can confirm that the high-velocity flow is driven near the protostar in Figures~\ref{fig:1}{\it g} and {\it h}. 
The flow speed exceeds $20\km$ near the protostar. 
In Figures~\ref{fig:1}{\it a}  and {\it b}, we can also confirm a thin pseudo disc  that surrounds the rotationally supported disc  around the protostar, 
 as also seen in observations \citep{riaz19,lee19}.
The outflow reaches $\sim500$\,au by this epoch. 
The knotty structure seen in the high velocity component (Figs.~\ref{fig:1}{\it a} and {\it b}) is  caused by episodic mass ejection \citep{machida14b,machida19}. 
The pseudo disc, rotationally supported disc  and low- and high-velocity outflows during  the early stage of the star formation are typically seen in both simulations and observations.

Figure~\ref{fig:2} shows three-dimensional views of the magnetic field lines, outflows and disc  for model T00 with different spatial scales. 
In the large scale, we can confirm a clear hourglass structure of the magnetic field lines (Fig.~\ref{fig:2}{\it a}).
The hourglass configuration has also been confirmed in observations \citep{girart06,girart09}. 
As the cloud collapses, the magnetic field lines are pulled toward the center of the cloud. 
Then, the magnetic field lines converge to the center and exhibit a wipe opening angle, as seen in Figures~\ref{fig:2}{\it b} and {\it c}. 
Inside the pseudo disc, there is a rotationally supported disc  that generates a strong toroidal field (Fig.~\ref{fig:2}{\it d}), because the magnetic field lines are twisted by the rotation motion of the  disc.
Both the twisted amplified  field and disc  rotation drive the outflow \citep{blandford82,uchida85}, as shown in Figure~\ref{fig:2}{\it d}, {\it e} and {\it f}.
Near the protostar, the configuration of the magnetic field lines and the structure of the outflows are very complex (Figs.~\ref{fig:2}{\it e} and {\it f}).  
However, it is clear that the net magnetic vector is parallel to the $z$-direction or the initial direction of the magnetic field at any scale.
In addition, the disc  normal and the outflow are directed in the $z$-direction. 

For this model (model T00),  the angular vector is aligned with the magnetic vectors in the initial cloud. 
The rotation and magnetic field are anisotropic forces and break the spherical symmetry in the collapsing cloud.
However, the alignment between the angular momentum and global magnetic field uniquely determines the direction of objects such as the pseudo disc, rotating disc, outflows and magnetic field at any scale. 
The alignment of the disc  normal, outflows and magnetic field evokes  a simple or classical picture of star formation.

\subsection{Rotation Axis Perpendicular to Magnetic Field: $\theta_0=90\dg$}
\label{sec:90}
The density and velocity distributions on the $x=0$, $y=0$ and $z=0$ cutting planes for model T90 are shown in Figure~\ref{fig:3}.
In the star formation process, two anisotropic forces (Lorentz  and centrifugal forces) produce a disc-like structure. 
The magnetic field produces a pseudo disc  \citep{galli93}, while the cloud rotation forms a rotationally supported (or Keplerian) disc  around a protostar.
The magnetic energy is larger than the rotational energy in the star-forming cloud \citep{crutcher99,caselli02}, and the magnetic field dissipates in the high density region of the collapsing cloud \citep{nakano02}.
Thus, the rotationally supported disc  appears at the small scale after the pseudo disc  forms at the large scale. 

As shown in Figures~\ref{fig:3}{\it b} and {\it e}, two nested discs appear in the star-forming cloud. 
Since the outer disc  corresponding  to the pseudo disc  forms along the magnetic field line,  its long axis is perpendicular to the global magnetic field. 
On the other hand, the inner disc  rapidly rotates, as shown in Figures~\ref{fig:3}{\it a}, {\it d} and {\it g}, and forms along the rotation axis. 
Thus, the long axis of the inner rotating disc  is perpendicular to the initial rotation axis and  parallel to the global magnetic field direction. 
For this model, we adopted the initial rotation axis as being perpendicular to the (uniform) global magnetic field (see Figs.~\ref{fig:3}{\it b} and {\it e}).
In this case, the magnetic field (or Lorentz force) cannot produce or reduce other components of the angular momentum,  because the initial magnetic direction is perfectly perpendicular to the rotation axis.
Therefore, the rotation direction maintains its initial direction, as seen in Figure~\ref{fig:3}. 
The cavity-like structure seen in Figure~\ref{fig:3} is caused by the generation of a toroidal field. 

The left panel of Figure~\ref{fig:4} shows a close-up view of Figure~\ref{fig:3}{\it h} 
and indicates that a weak outflow appears  and creates a density cavity near the  protostar.
Around the protostar, the magnetic field lines are strongly bent and twisted, because the circumstellar disc  is rapidly rotating. 
Thus, a strong toroidal field is generated, and the magnetic pressure gradient force drives the outflow.
The cavities or density gaps seen in Figures~\ref{fig:3}{\it b}, {\it c}, {\it e}, {\it f} and {\it h} are created by the amplified magnetic field.

The configuration of the magnetic field lines and the structure of the disc  for model T90 at  $t_{\rm ps} =509.7$\,yr are plotted in Figure~\ref{fig:5}.
At the large scale, although the whole structure of the magnetic field lines is slightly distorted, we can confirm an hourglass-like configuration in the magnetic field lines. 
A similar configuration of the magnetic field lines can be confirmed in observations \citep[e.g.][]{shinnaga12}.
In Figures~\ref{fig:5}{\it b} and {\it c}, we can infer  that a disc-like structure forms along the magnetic field lines, in which the disc  is gradually tilted.  
Thus,  the disc  is formed by the magnetic effect at the large scale of $\gtrsim 370$\,au.
 
At the middle scale (Figs.~\ref{fig:5}{\it d} and {\it e}), although coherent magnetic fields are realized, their direction differs from that of the large-scale field.  
Since the magnetic field dissipates and  magnetic braking weakens, a rotating disc  forms along the initial rotation axis at this scale ($\lesssim 100$\,au).
As the rotating disc  forms,  the weak magnetic fields are passively transformed and are crushed or pushed against the disc  long axis or along the disc  surface.
Then,  magnetic field lines along the disc  short axis appear,  as shown in Figures~\ref{fig:5}{\it d} and {\it e}.
At the small scale, a toroidal field is generated along the rotation axis of the circumstellar disc  due to the disc  rotation (Fig.~\ref{fig:5}{\it f}). 
This configuration of the magnetic field lines was thoroughly investigated in \citet{machida06}.

For model T90, the rotation axis is maintained in the $x$-direction, while the direction of the local magnetic field significantly varies at different scales.
To determine, or visualize, the directions of the magnetic field and angular momentum at different scales, we averaged the magnetic field and angular momentum within a radius (magnetic field) or in a high density region  (angular momentum) at each epoch, according to equations (1)--(6) of Paper I.
As described in Paper I, we calculated the magnetic field direction in a spherical volume within a radius $r_{\rm th}=10$, 100 and 100\,au, and averaged it by the volume  (see  eq.~[6] of Paper I).
The angular momentum direction was calculated and averaged by the mass in a region of $n > n_{\rm th}$, in which threshold densities of $n_{\rm th}=10^7$, $10^{10}$ and $10^{13}\cm$ are  adopted (see eq.~[5] of paper I). 

Figure~\ref{fig:6} plots a time series of the directions of the magnetic field (left) and angular momentum (right) during $\sim500$\,yr after protostar formation.
The figure indicates that the directions of the magnetic field at the small scale differ considerably from the initial direction (i.e. the $z$-axis). 
In addition, the direction of the magnetic field varies with time, while the rotation direction is maintained in the $x$-direction (i.e. initial direction). 
The disc  forms along the magnetic field lines at the large scale, while it forms along the rotation axis at the small scale.
Thus, the disc  normal is determined by the balance between the Lorentz and centrifugal forces, and hence the direction of the disc  normal varies according to the spatial scale.
In addition, the disc  normal should change with time. 
In the following subsections, we show the general case for $0\dg < \theta < 90\dg$.

\subsection{Rotation Axis Misaligned with Magnetic Field: $ \theta_0 = 30\dg$}
Figure~\ref{fig:7} shows the density and velocity distributions on the $x=0$, $y=0$ and $z=0$ planes for model T30.
In the figure, we can confirm that the outflow is strongly ejected  by the rotating disc.
At the large scale, we can see that a filamentary structure is formed from a pseudo disc, in which the pseudo disc  is twisted by the rotation motion \citep{takahashi19}. 
As a whole,  the structures seen in this model in Fig.~\ref{fig:7} are basically the same as for model T00 (Fig.~\ref{fig:1}).
In both models, the outflow, a (twisted) pseudo disc and a rotationally supported disc  appear.
The most significant difference between models T00 and T30 is the directions of the outflows and disc.
These directions in the misalignment model T30 significantly change at different spatial scales (Fig.~\ref{fig:7}), while the directions do not change in the aligned model T00 (Fig.~\ref{fig:1}).
In the misalignment model, the directions seem to be significantly changed, even for different viewing angles.

Figure~\ref{fig:8} shows a three-dimensional view of the magnetic field lines, outflows and disc  at $t_{\rm ps}=670.7$\,yr for model T30. 
Although the magnetic field lines have an hourglass-like configuration at the large scale (Figs.~\ref{fig:8}{\it a} and {\it b}),  they have a very complicated configuration at the small scale (Fig.~\ref{fig:8}{\it c}--{\it f}). 
We can also confirm that the disc  normal direction differs at each scale. 
In addition, the outflow direction changes dependent on its velocity, as already shown and discussed  in Paper I.
In Figures~\ref{fig:8}{\it c} and \ref{fig:8}{\it d}, the low-velocity outflow components are coloured blue and green, while the high-velocity component is coloured yellow.
The panels indicate that the high-velocity components are tilted toward the low-velocity component.
As seen in Figures~\ref{fig:8}{\it c} and \ref{fig:8}{\it d}, the angle between the $z$-axis and the outflow direction is smaller in the low-velocity component than in the high-velocity component (for details, see Paper I).
Thus, the different directions of outflows are nested. 
Recent ALMA observation also confirmed a different direction for the flows at a very early stage of star formation \citep{matsushita19}.

Near the protostar, the magnetic field lines are strongly twisted by the rotation motion of the disc  (Fig.~\ref{fig:8}{\it f}).
In addition, the outflow directions are considerably different from the initial direction of the magnetic field.
Also, the disc  normal is different from the initial direction of the angular momentum (i.e., the $x$-direction).
Thus, at the middle and small scales ($\lesssim 1000$\,au), the magnetic field lines, outflows and disc  have no specific information about the initial (or large-scaled) directions of magnetic field and angular momentum (for details, see \S\ref{sec:directions}).

\subsection{Other Misalignment Models}
In this subsection, we briefly comment on the four models with $\theta_0=5$, 10, 60 and 80$\dg$ (models T05, T10, T60 and T80).
These models have interesting features, some of which seem to reproduce specific observations. 

Figure~\ref{fig:9} shows the density and velocity distributions on the $y=0$ plane for the models T05, T10, T60 and T80.
For model T05 (Figs.~\ref{fig:9}{\it a}--{\it c}), the initial small angle difference results in a noticeable tilt of the disc  normal and outflow directions toward the $z$-axis.
In Figure~\ref{fig:9}{\it b}, we can see a warped pseudo disc. 
In addition, the angle difference between the initial magnetic field and angular momentum produces a knotty jet-like configuration (Fig.~\ref{fig:9}{\it c}), which is often seen in observations \citep[e.g.][]{Cunningham09}.
The speed of the jet-like flow exceeds $20\km$. 

When the angle difference is as small as $\theta_0=10\dg$ (model T10), the disc  around the protostar is highly distorted, as shown in Figure~\ref{fig:9}{\it d}. 
In this model, the disc  normal and outflow directions rapidly change with time (for details, see \S\ref{sec:directions}). 
Therefore, the outflow has a wide opening angle  (Fig.~\ref{fig:9}{\it e} and {\it f}).
Wide-angle outflows are also seen in some observations \citep[e.g.][]{velusamy98}.

In model T60, we can confirm a strong mass ejection. 
In Figure~\ref{fig:10}{\it i}, high-density clumps are embedded in the outflow. 
The inclined outflow interacts with the pseudo disc  and strips its surface.
The stripped gas from the pseudo disc  is ejected as a part of the outflow. 
The interaction between the outflow and pseudo disc  produces some gas clumps in the outflow, which is also seen in observations \citep{plunkett13,osorio17}.

The outflow direction drastically changes over time in  model T80. 
In Figure~\ref{fig:9}{\it l}, we can see a quadrupole-like shock, which is caused by outflows with different directions.
This kind of quadrupolar or multipolar outflow is confirmed in observations \citep{gerin15}. 
In this model, the outflow propagates roughly along the $x$-axis at early times. 
Then, the outflow changes direction and  propagates mainly close to the $z$-direction.
As a result, a quadrupolar cavity is formed.

Figure~\ref{fig:10} plots three-dimensional views of the magnetic field lines, high-density region and outflow for model T80.
As seen in Figure~\ref{fig:10}{\it b}, the ejected knot changes the local configuration of the magnetic field lines. 
Although the outflow is driven by the magnetic effect,  the magnetic field is also affected by the outflow.
In Figure~\ref{fig:10}{\it c}, a quadrupolar outflow is indicated by arrows. 
When the initial rotation axis is highly inclined from the global field, the disc  normal and outflow directions rapidly change over time (see, \S\ref{sec:directions}).

As shown in Figure~\ref{fig:9}, the initial angle difference produces various morphologies for the outflows.
Thus, the angle difference could explain the different morphologies seen in observations \citep[e.g.][]{velusamy14}.

\subsection{Directions of Magnetic Field, Outflow and Disc }
\label{sec:directions}
The structures of the magnetic field, outflow and disc  at $t_{\rm ps}\simeq500$\,yr for all the models are summarized in Figure~\ref{fig:11}, in which the spatial scale and viewing angle are the same in all the panels.
In all the models, the initial magnetic field is set to be parallel to the $z$-direction.
Thus, the initial direction of the magnetic field is upward in Figure~\ref{fig:11}. 
At the scale of $\sim1000$\,au, we can confirm an hourglass-like configuration of the magnetic field in all the models.
However, except for model T00, the net magnetic field directions seem to be inclined from the initial direction (i.e. $z$-direction), especially for models T30, T45, T60, T80, T85 and T90.
The net magnetic field directions gradually change from the initial direction as the spatial scale decreases.

The figure also shows that the outflow directions differ in each model and they are not aligned with the initial  direction of the magnetic field (i.e. the $z$-direction), except for model T00.
The outflow direction is roughly parallel to the $z$-axis for models with small $\theta_0$, T00, T05 and T10.
On the other hand, the outflow directions are considerably inclined  from the initial magnetic field direction for models with large $\theta_0$, T30, T45, T60, T80 and T85. 
In addition, in the models T10, T30, T45, T60 and T80, the outflow directions differ in every velocity component.
For example, in model T10, the low-velocity component (blue surface; $v_r\simeq3\km$) is roughly parallel to the $z$-axis, while the high-velocity component (green or orange surface; $v_r\sim35\km$) is substantially inclined from the $z$-axis. 
Thus, the propagation directions of the outflows differ in every velocity component.
Therefore, in addition to the scale of the outflow, we need to pay attention to the outflow velocity to identify the outflow direction. 

Next, in Figure~\ref{fig:11} we provide an overview of the directions of the disc-like structure (red iso-density surface at the center)  and outflow (blue, green and orange iso-velocity surfaces). 
Note that in Figure~\ref{fig:11}, the surface density of the disc-like structure is set to be $3.9\times10^8\cm$.
Thus, the disc  seen in the figure corresponds to a pseudo disc, not a rotationally supported disc  \citep{machida06}.
The directions of the rotationally supported disc  are shown in Figure~\ref{fig:13}.
In Figure~\ref{fig:11}, the outflow is aligned with the disc  normal only for model T00 (Fig.~\ref{fig:11}{\it a}).
In models T10, T30, T45 and T60, the relative angle between the disc  normal and the outflow seems to be about $90\dg$.
Thus, the outflow direction is roughly perpendicular to the disc  normal at the scale of $\sim1000$\,au.
For model T85, the outflow seems to be aligned with the disc  normal. 
These results shows that, at this scale, there is no very clear correlation between the magnetic field, outflow and disc  normal directions.

To determine the time variation of the directions of the magnetic field, disc  (or angular momentum) and outflows, the directions at each time are plotted in Figure~\ref{fig:12}.
Note that, in Paper I, we showed that the disc  normal $\vect{n}$ is almost parallel to the angular momentum $\vect{J}$ in the high density region $n \gtrsim 10^{10}\cm$, above which the outflow emerges \citep{machida06,machida07,machida13}.
Thus, to avoid complexity, we only plotted the directions of the angular momentum  instead of the disc  normal.
As described above, the derivation of the directions are fully described in Paper I.
A good agreement between the disc  normal and angular momentum in the high density region was confirmed in Paper I. 

The directions of the magnetic field at different scales are plotted in the left column of Figure~\ref{fig:12}.
In each model, the direction of the net magnetic field at a scale of $1000$\,au roughly maintains its initial direction, i.e. the $z$-direction (see the blue dots in the left column).
On the other hand, the net magnetic field at the scale of $100$\,au (green dots)  is significantly  inclined from the initial direction.
Furthermore, at the scale of $10$\,au (red dots), the magnetic field direction significantly changes over time, and considerably differs from the initial direction.
Thus,  the magnetic fields at the small scale of $\lesssim 100$--$1000$\,au do not retain their initial or large-scaled directions.

The directions of the angular momentum are plotted in the middle column of Figure~\ref{fig:12}.  
As described in \S\ref{sec:settings}, the rotation axis is inclined from the $z$-axis to the $x$-axis by $\theta_0$.
The rotationally supported disc  forms in the region of $\gtrsim10^{10}\cm$ where the magnetic field dissipates and magnetic braking becomes ineffective \citep{machida11, tomida15,tsukamoto15}.
The discs with a high density of $\ge10^{10}\cm$ (green dots) and $\ge10^{13}\cm$ (red dots) seem to keep the initial direction of the angular momentum while staying around the initial direction, 
which results in oscillation of the disc  surface where the protostellar outflow emerges.

The structure of the rotationally supported disc  at $t_{\rm ps}\simeq500$\,yr for all models is plotted in Figure~\ref{fig:13}.
The figure indicates that the disc  gradually inclines from the $z$-direction as the initial mutual angle $\theta_0$ increases. 
A comparison of Figure~\ref{fig:11} and Figure~\ref{fig:13} indicates that, in each model, the outflow direction near its driving region is roughly aligned with the disc  normal. 
Note that the viewing angle is the same in Figures~\ref{fig:11} and \ref{fig:13}.
Thus, the outflow emerges along the local disc  normal. 
The disc  normal direction differs at every scale and for every density.
In the middle column of Figure~\ref{fig:12}, the direction of the angular momentum at the low density of $n_{\rm th} =10^{10}\cm$ is slightly different from that at the high density of $n_{\rm th}=10^{13}\cm$, which indicates that the disc  normal differs at every scale. 
However, the difference in the angles is not very large.  
A recent ALMA observation has revealed the misalignment rotation axis or the disc  normal  \citep{sakai19}.
 
The disc  normal direction also changes with time, indicating that the disc  normal oscillates. 
The oscillation of the rotation axis results in different directions of the outflow. 
In Figure~\ref{fig:12}, we can confirm that the disc  normal direction nearly  corresponds to the outflow directions. 
Roughly, the angular momentum direction of the low-density disc  region corresponds to the low-velocity outflow, while that of the high density disc  region corresponds to the high-velocity flow, as discussed in \citet{matsushita19}.

\section{Discussion}
\subsection{Mass of Protostar, Disc  and Outflow, and  Outflow Momentum} 
Many past  studies have investigated the star formation process  assuming that the initial rotation axis is parallel to the global magnetic field \citep{tomisaka02,banerjee06,commercon11,seifried12,tomida13,machida14}. 
These studies could well reproduce the objects and phenomena observed in star-forming regions, such as pseudo discs, rotationally supported discs and outflows. 
In this study, we assumed that the rotation angle is inclined toward the global magnetic field.
Our simulation could also reproduce  pseudo discs, rotationally supported discs  and outflows, 
showing that when the initial rotation axis is not aligned with the global magnetic field, the star formation process does not change qualitatively. 
On the other hand, our results showed that it is important to pay attention to vector quantities, such as angular momentum and outflow momentum, as their directions change over time.
In this section, we compare some physical quantities (or scalar quantities) between the models. 

Figure~\ref{fig:14}{\it a} shows the time evolution of the protostellar mass.
The protostellar mass is estimated as 
\begin{equation}
M_{\rm ps} = \int_{n>10^{18}\cm \ {\rm and} \ v_{r}<1\km} \rho\, dv,
\end{equation}
where the condition $v_{r}<1\km$ is imposed so as to exclude the high-density outflow component \citep{machida19}. 
The figure indicates that the initial angle difference does not significantly affect the determination of the protostellar mass. 
The protostellar mass for all models is in the range of $0.01 \msun <M_{\rm ps}<0.02 \msun$ at $t_{\rm ps}=30$\,yr and  $0.02\msun <M_{\rm ps}<0.04\msun$ at $t_{\rm ps}=500$\,yr.
Thus, the mass accretion rate is estimated to be $\sim 4$--$8\times10^{-5}\msun$\,yr$^{-1}$.
The difference in the mass and mass accretion rate among the models is within a factor of two. 

Figure~\ref{fig:14}{\it b} plots the disc  mass for all models against the elapsed time after protostar formation.
The disc  mass is estimated according to the procedure described in Paper I. 
Unlike the protostellar mass, there is over one order of magnitude difference in the disc  mass. 
At $t_{\rm ps}\sim500$\,yr, the disc  mass for model T80 is $1.6\times10^{-3}\msun$, while that for model T05 is $5.0\times10^{-2}\msun$. 
The disc  mass is smaller in models with lager $\theta_0$ than in models with smaller $\theta_0$. 
However, the disc  mass for model T90 rapidly increases  with time, and roughly catches up with the disc  mass of model T00 at $t_{\rm ps} \sim 500$\,yr.
Although further long-term simulations are necessary to determine the disc  mass, the diversity of the disc  mass may lead to the diversity of planetary systems.
A detailed discussion of the angle dependence  $\theta_0$ on the disc  mass and radius is described  in \citet{hirano19b}.

The outflow mass and outflow momentum are plotted in Figure~\ref{fig:14}{\it c} and {\it d}, respectively. 
They are estimated as 
\begin{equation}
M_{\rm out} = \int_{v_r > 1 \km} \rho \, dV,
\end{equation} 
\begin{equation}
P_{\rm out} = \int_{v_r>1\km} \rho v \,dv,
\end{equation}
where $v_r$ is the radial velocity and $v$ is defined as $v=(v_x^2+v_y^2+v_z^2)^{1/2}$. 
As shown in \S\ref{sec:90}, a very weak outflow appears in model T90. 
Thus, the outflow mass and momentum in model T90 are significantly smaller than in other models. 
Furthermore, among the models with $\theta_0<90\dg$, the models having larger angles, models T45, T60 and T80, tend to have larger outflow mass and momentum. 
As shown in Figure~\ref{fig:9}, in such models, the outflow interacts with the pseudo disc  and strips the gas of the pseudo disc.
Therefore, such models can have a large outflow mass and momentum. 
Although there exists a difference in the outflow mass and momentum, the difference between the models is within one order of magnitude at $t_{\rm ps}\simeq500$\,yr except for model T90. 
Thus, the initial angle difference does not cause a significant difference in the outflow physical quantities.

\subsection{Disc  Direction}
In this study, we adopted a strong magnetic field, as described in \S\ref{sec:settings}.
Thus, the angular momentum perpendicular to the magnetic field should be preferentially transferred \citep{mouschovias79,mouschovias80,mouschovias80b}. 
Interestingly however, the initial rotation axis seems to be roughly preserved (Fig.~\ref{fig:13}). 
In the star-forming cloud, since the gas rapidly collapses in the early phase (or the low-density region), magnetic braking is not fully effective \citep{machida11,tsukamoto18}.
On the other hand, in the high-density region, the magnetic field dissipates and magnetic braking becomes ineffective.
Thus, a rotationally supported disc forms almost along the initial rotation axis \citep{machida14b,hirano19}.

After disc  formation,  the angular momentum of the disc  is transferred by the outflow and magnetic braking.
The disc  first forms in the magnetically inactive region \citep{machida14,machida16}; the magnetic field lines are passively bent roughly along the disc  normal direction. 
With such a magnetic configuration, the magnetic field  transfers the angular momentum preferentially perpendicular to the disc  normal \citep{matsumoto04}.
As a result,  although the disc  normal oscillates, it thereby maintains the initial direction during the main accretion stage.

\subsection{Outflow Mass and Velocity}
As described in \S\ref{sec:intro}, there exists a discrepancy in the outflow velocity between  simulations and observations. 
The outflow velocities calculated in many past simulations are within $\sim\km$ \citep[][and references therein]{inutsuka12}, while those in observations are $\sim10$--$100\km$ \citep{arce07}. 
In many simulations, the region around the protostar is masked by sink cells or sink particles.
Thus, the high-velocity component never appears in such simulations, because it is driven near the protostellar surface \citep{machida06b}. 
On the other hand, since this study resolved the protostar itself, the high velocity component appears. 

Figure~\ref{fig:15} plots the outflow mass against the outflow radial velocity $v_{\rm rad}$ at the end of the simulation, in which the outflow mass at every velocity is estimated as
\begin{equation}
M_{\rm enc}(v_{\rm rad}) = \int_{v_{\rm rad}}^{v_{\rm rad}+\delta v} \rho \ dV,
\end{equation}
where $\delta v=0.05\km$ is adopted.
As shown in the figure, although the outflow mass is small, the outflow velocity exceeds $\gg20\km$ except for model T90.
In addition, except for model T90, the maximum outflow velocity during the simulation exceeds $\sim50$--$100\km$. 
A detailed analysis of the outflow and its velocity is beyond the scope of this study, suffice it to note that the outflow velocities in this simulation are comparable to observations.

\section{Summary}
As described in \S\ref{sec:intro}, the star formation process has been investigated in simple settings, in which the initial rotation axis is assumed to be parallel to the global magnetic field in simulations. 
In this case, the directions of outflow, disc  normal and magnetic field lines at any scale are aligned. 
On the other hand, \citet{hull13} pointed out that the outflow directions are not aligned with global magnetic fields. 
Thus, past numerical simulations seem not to agree with the observations. 
Some researchers have pointed out that the outflow direction is not always aligned with the global magnetic field, when the initial rotation axis is inclined from the global magnetic field \citep{matsumoto04}. 
However, in the past simulations, since the protostar is not resolved with sufficient spatial resolution, the speed of outflow is  $<10\km$, which is significantly slower than that measured in observations.
In our previous study (Paper I), we showed that the directions of outflow, disc  normal and magnetic field lines are not aligned in the early star formation stage, during  which the outflow speed exceeds $\gtrsim 50$--$100\km$ \citep[see also][]{machida19}. 
However, in Paper I, we only presented a single model ($\theta_0=45\dg$).
Thus, we do not know whether or not the angle difference is generally maintained. 

In this study, we investigated the evolution of clouds having different mutual angles between the initial cloud rotation  axis and global magnetic field. 
The simulations showed that  the directions of outflow, disc  and local magnetic field are never aligned during the early accretion stage. 
In summary, there are no clear correlations in angles between the outflow, local magnetic field and disc. 

The disc  normal is determined by the balance between the centrifugal and Lorentz forces. 
In star-forming clouds, since the rotation degree and magnetic field strength differ at different spatial scales and in different epochs, the disc  normal changes accordingly. 
In addition, since the protostellar outflow is driven by the rotation and magnetic field around the disc, the outflow directions are not simply determined. 
Furthermore, the outflow itself changes the configuration of the magnetic field, because the magnetic field lines far from the outflow driving region are bent by the ejected clumps.  
The configuration of the magnetic field is closely related to the magnetic braking and thus the efficiency of the angular momentum transfer, which changes the  normal direction of the rotationally supported disc. 
Thus, many factors are related to determining the directions. 

The past star formation simulations assuming alignment between the cloud rotation and magnetic field  are not unrealistic. 
Although we should not compare such simulations with observations, especially when interpreting the directions of outflows and magnetic field,  the star formation process constructed from the aligned setting is not wrong. 
Also, in the misalignment cases, the pseudo disc, rotationally supported disc, and low- and high-velocity outflows appear, as in the aligned cases. 
In addition, the outflow masses and momentum between the aligned and misalignment cases are not significantly different.
The only difference between the aligned and misalignment cases is the directions of the outflow, disc  and magnetic field lines. 
The different directions somewhat change the star forming environment at a scale of $\lesssim 1000$\,au, and give rise to complex structures, as seen in observations.
At present, we can fairly compare simulations and observations only in the very early phase of star formation.
Since the probability of the existence of such early-phase objects is low, we require further long-term calculations to compare star formation between simulations and observations for all phases.

\section*{Acknowledgements}
The authors would like to thank B.~ Riaz, K.~ Tomida and S.~ Takasao for their helpful contributions.
We also thank the referee R.~Banerjee for careful reading and useful comment on this paper. 
The present research used the computational resources of the HPCI system provided by  Cyber Science center, Tohoku University and  Cybermedia Center, Osaka University through the HPCI System Research Project (Project ID: hp170047, hp180001, hp190035). The simulations reported in this paper were also performed by the 2018 and 2019 Koubo Kadai on Earth Simulator (NEC SX-ACE) at JAMSTEC.
This work was supported by a JSPS Research Fellowship for SH and JSPS KAKENHI Grant Numbers 18J01296 for SH and by 17K05387, 17H06360, and 17H02869, 17KK0096 for MNM.


\clearpage

\renewcommand{\arraystretch}{1.2}
\begin{table}
\begin{center}
\begin{tabular}{c||cccccccc} \hline
Model & $\theta_0$ & $B_0$ [G] & $\Omega_0$ [s$^{-1}$] & $\alpha_0$ & $\beta_0$ & $\mu_0$ & $t_{\rm ps, end}$\,[yr]  & $M_{\rm ps, end}$\, [$\msun$] \\
\hline
T00 & 0$\dg$ & \multirow{9}{*}{$5.7\times10^{-5}$}  & \multirow{9}{*}{$1.4\times10^{-13}$} & \multirow{9}{*}{0.39}  &  \multirow{9}{*}{0.026}& \multirow{9}{*}{1.2} & 674& 0.034 \\
T05 & 5$\dg$ & & & & & & 418& 0.021  \\
T10 & 10$\dg$ & & & & & & 786& 0.045 \\
T30 & 30$\dg$ & & & & & & 671& 0.043 \\
T45 & 45$\dg$ & & & & & & 573& 0.039 \\
T60 & 60$\dg$ & & & & & & 573& 0.033 \\
T80 & 80$\dg$ & & & & & & 682& 0.025 \\
T85 & 85$\dg$ & & & & & & 561& 0.040 \\
T90 & 90$\dg$ & & & & & & 510& 0.032 \\
\hline
\end{tabular}
\end{center}
\caption{
Model name, initial cloud parameters and calculation results.
Column 1 gives the model name. 
Column 2 gives the parameter $\theta_0$. 
Columns 3 and 4 give the magnetic field strength $B_0$ and angular velocity $\Omega_0$ at the initial state. 
Columns 5 and 6 give the ratios of the thermal $\alpha_0$ and rotational $\beta_0$ energies to the gravitational energy of the initial cloud. 
Column 7 gives the initial mass-to-flux ratio $\mu_0$ normalized by the critical value.
Columns 5 and 6 give the elapsed time after protostar formation $t_{\rm ps, end}$ and protostellar mass $M_{\rm ps, end}$ at the end  of each simulation.
}
\label{table:1}
\end{table}
\begin{figure*}
\begin{center}
\includegraphics[width=1.0\columnwidth]{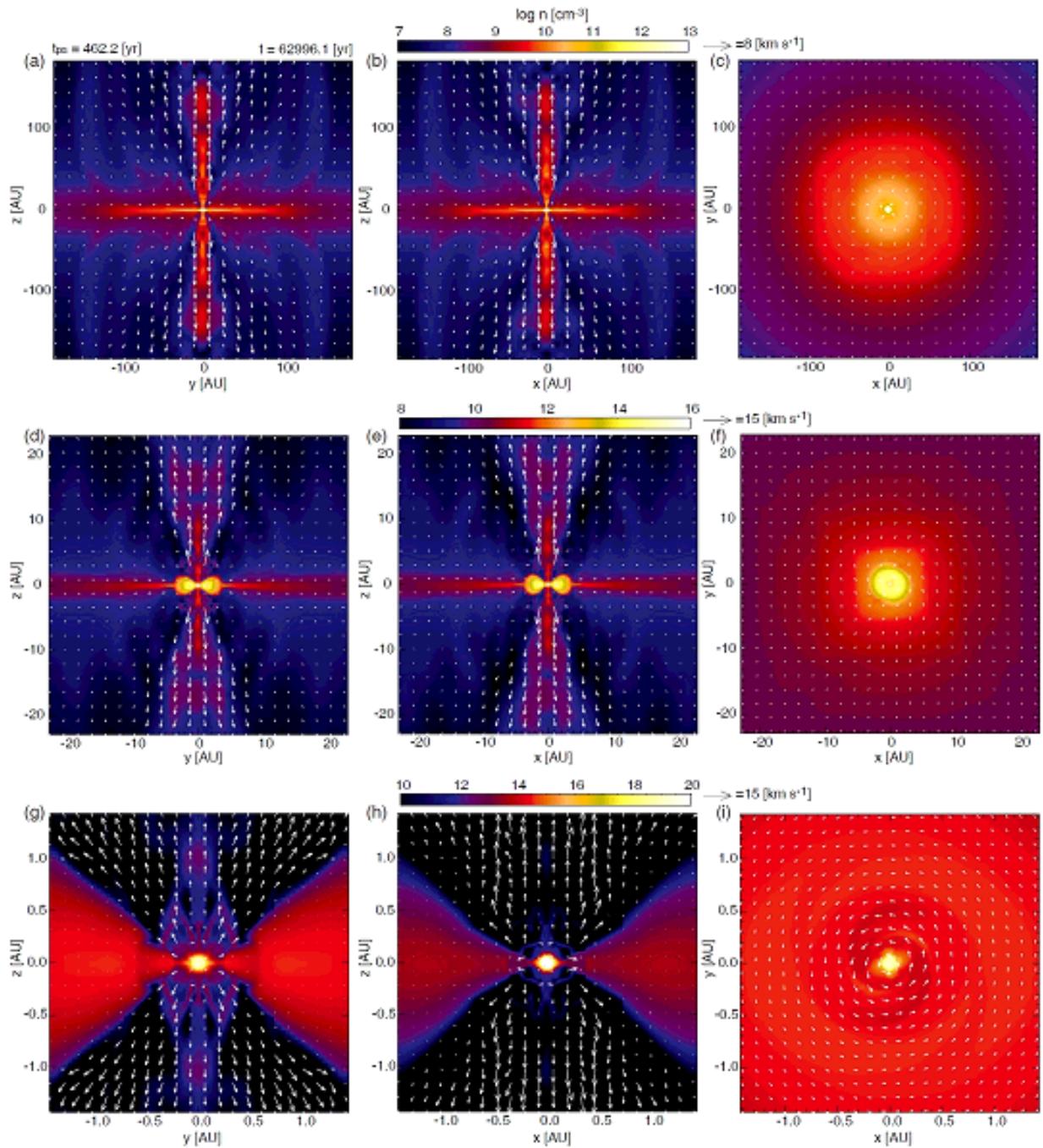}
\end{center}
\caption{
Density (colour) and velocity (arrows) distributions on the $x=0$ (left column), $y=0$ (middle column) and $z=0$ (right column) planes for model T00 with a box size of 295\,au (top panels), 37\,au (middle panels) and 2.3\,au (bottom panels).
The elapsed time after protostar formation $t_{\rm ps}$ and that after the beginning of the cloud collapse $t$ are described in the upper part of panel ({\it a}). 
}
\label{fig:1}
\end{figure*}
\clearpage

\begin{figure*}
\begin{center}
\includegraphics[width=1.0\columnwidth]{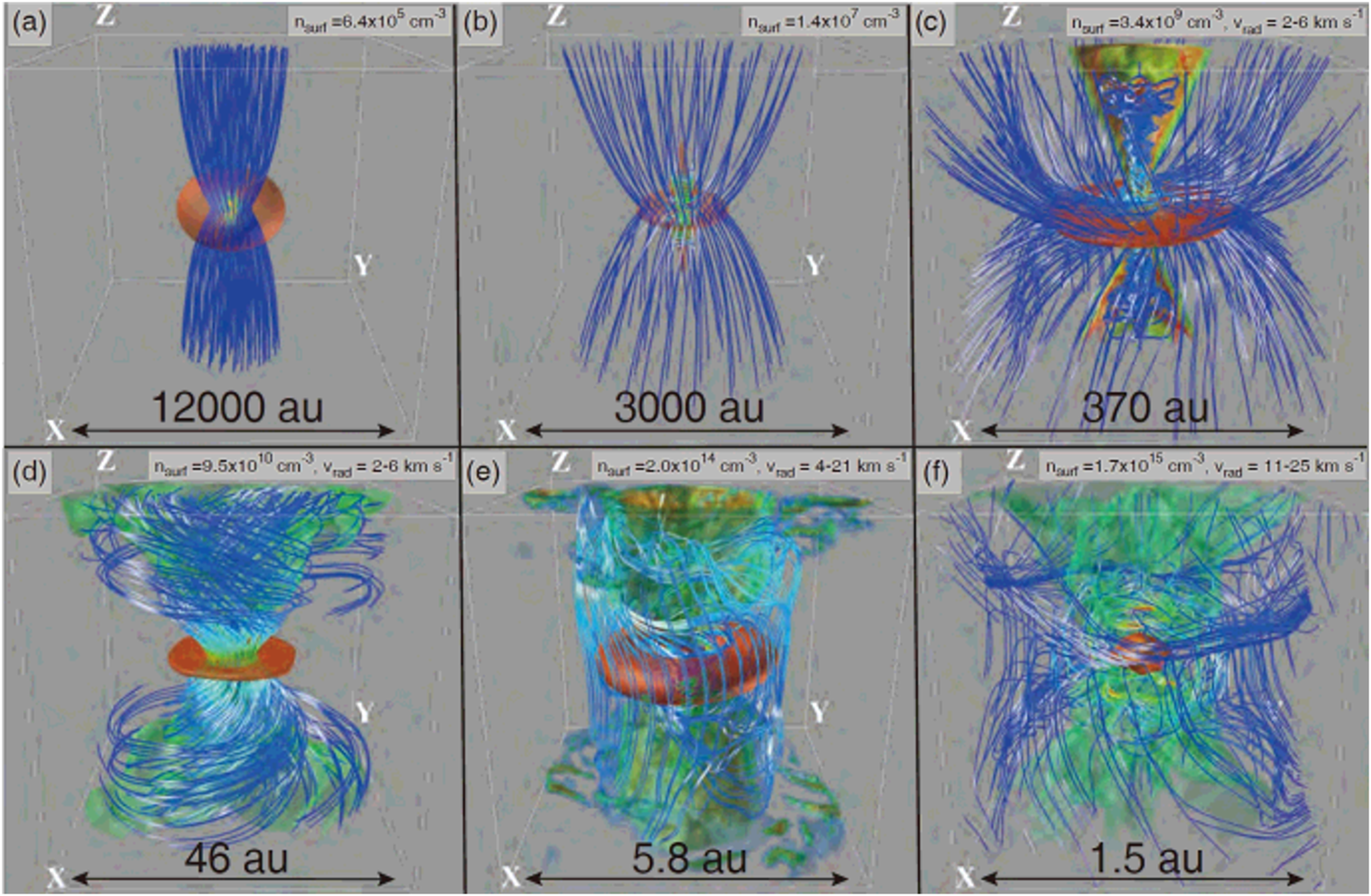}
\end{center}
\caption{
Three-dimensional view of magnetic field lines (blue lines), high-density region (orange iso-density surface) and outflows (blue, green and yellow iso-velocity surfaces) at $t_{\rm ps}=462.2$\,yr and $t=62996.1$\,yr with different spatial scales for model T00.
The box size is described in each panel.
The density of the iso-density surface $n_{\rm surf}$ and the velocity range of the outflow $v_{\rm rad}$ are also described in each panel. 
The viewing angle is the same in all panels. 
}
\label{fig:2}
\end{figure*}
\clearpage

\begin{figure*}
\begin{center}
\includegraphics[width=1.0\columnwidth]{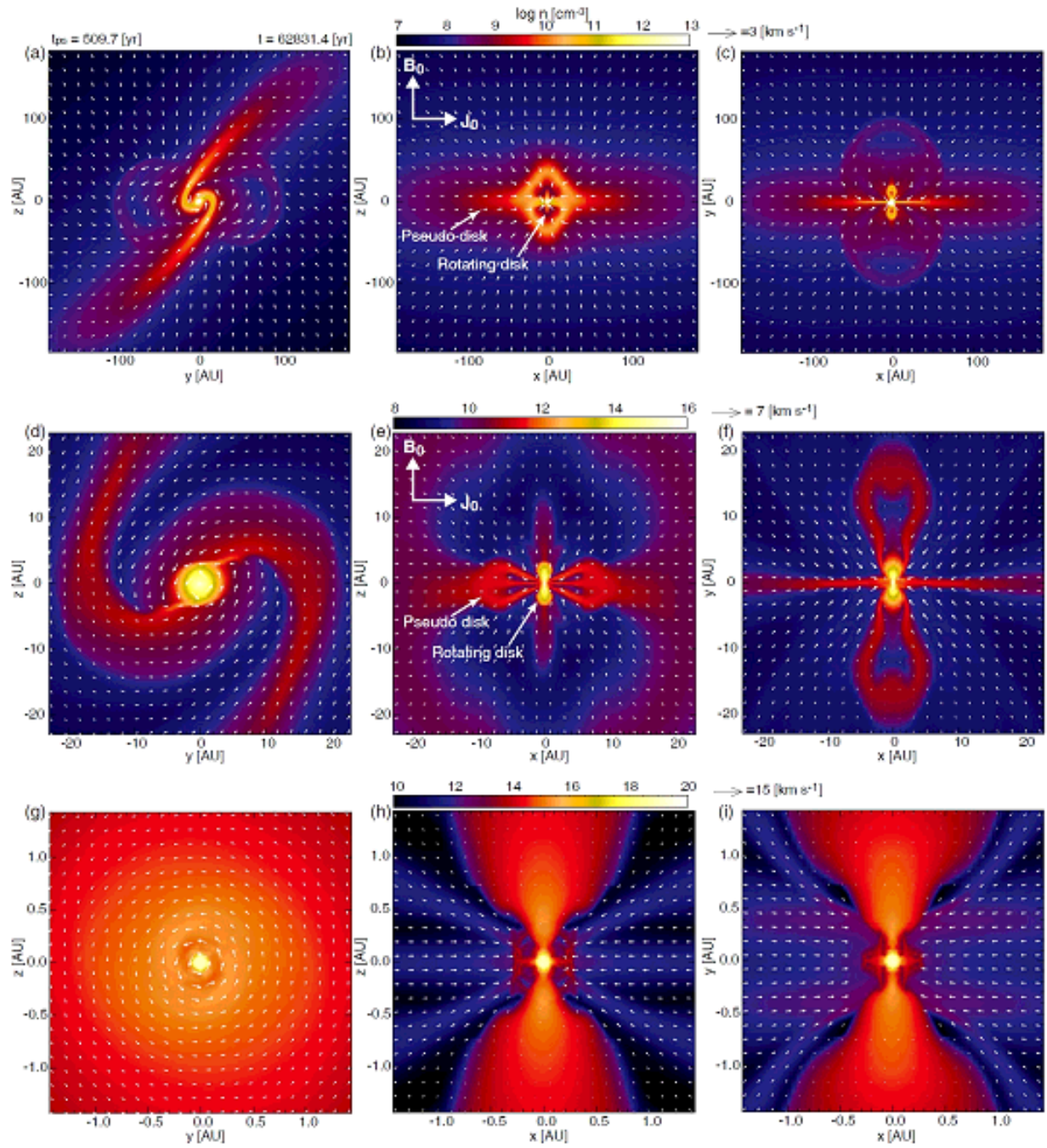}
\end{center}
\caption{
As for Figure~\ref{fig:1}, but for model T90. 
The initial direction of the magnetic field and rotation vector are plotted in panels ({\it b}) and ({\it e}).
The pseudo disc  and rotating  disc  are labeled in panels ({\it b}) and ({\it e}).
}
\label{fig:3}
\end{figure*}
\clearpage

\begin{figure*}
\begin{center}
\includegraphics[width=1.0\columnwidth]{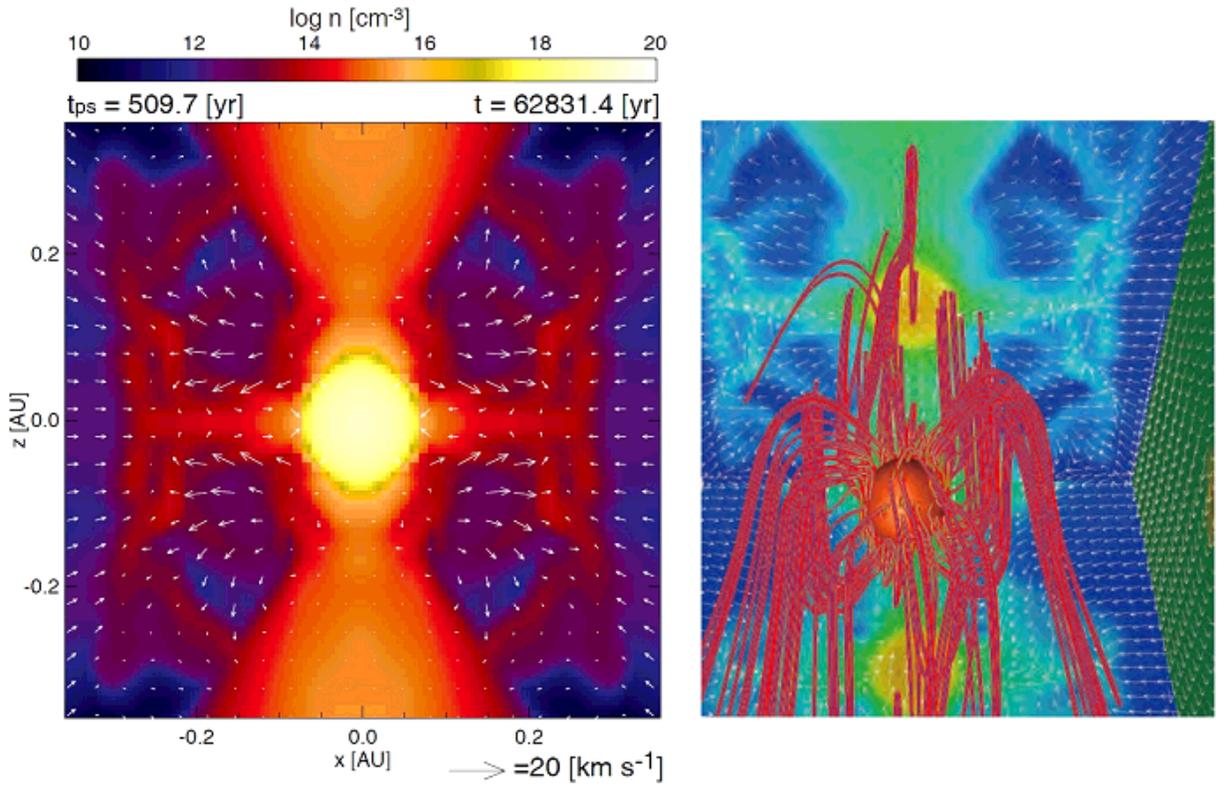}
\end{center}
\caption{
Left: Density (colour) and velocity (arrows) distributions on the $y=0$ plane for model T90. 
The elapsed time after protostar formation $t_{\rm ps}$ and that after the beginning of the cloud collapse $t$ are described in the upper part.
Right: Three-dimensional view of magnetic field lines (red lines) and protostar (orange surface) at the same epoch as in the left panel. 
The density and velocity distributions on the $x=0$, $y=0$ and $z=0$ plane are projected  on each wall surface.
}
\label{fig:4}
\end{figure*}
\clearpage

\clearpage
\begin{figure*}
\begin{center}
\includegraphics[width=1.0\columnwidth]{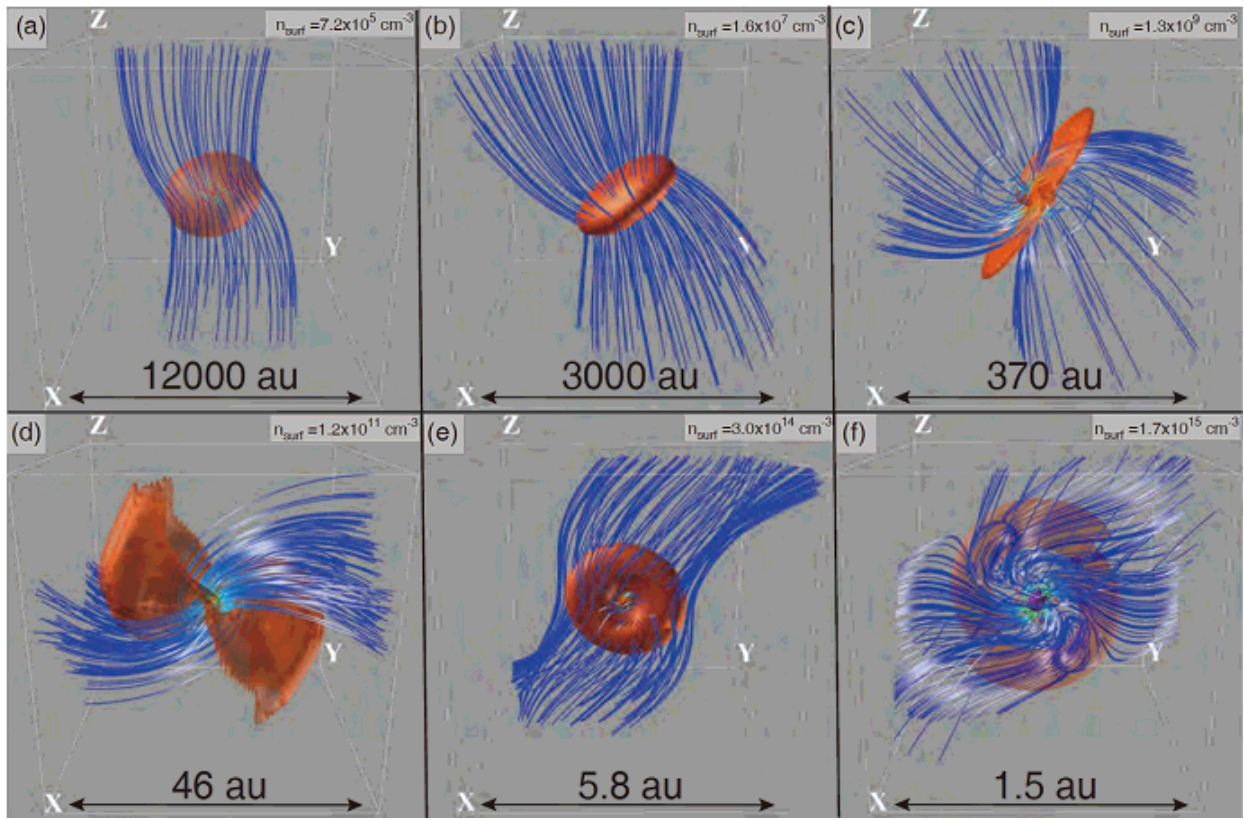}
\end{center}
\caption{
As for Figure~\ref{fig:2}, but for model T90.
}
\label{fig:5}
\end{figure*}
\clearpage

\begin{figure*}
\begin{center}
\includegraphics[width=1.0\columnwidth]{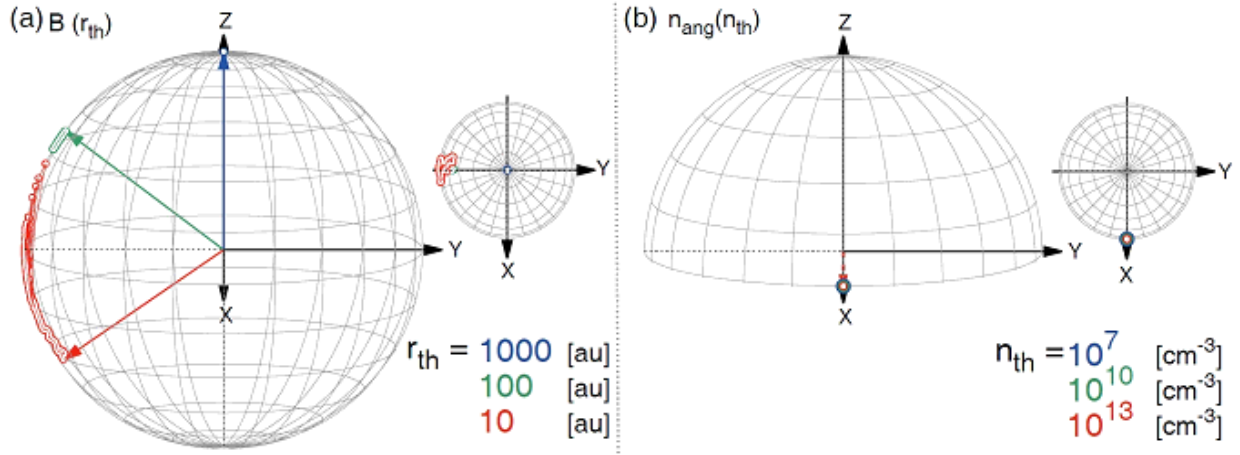}
\end{center}
\caption{
Time evolution of the directions of the magnetic field plotted on a spherical surface (a) and angular momentum plotted on the surface of a hemisphere (b).
The magnetic field directions are measured at three different scales of 1000, 100 and 10\,au.
The angular vectors are measured  for three different threshold densities of $n_{\rm th}=10^7$, $10^{10}$ and $10^{13}\cm$.
The arrows indicate the direction at the end of the simulation ($t_{\rm ps}$=510\,yr).
}
\label{fig:6}
\end{figure*}
\clearpage

\clearpage
\begin{figure*}
\begin{center}
\includegraphics[width=1.0\columnwidth]{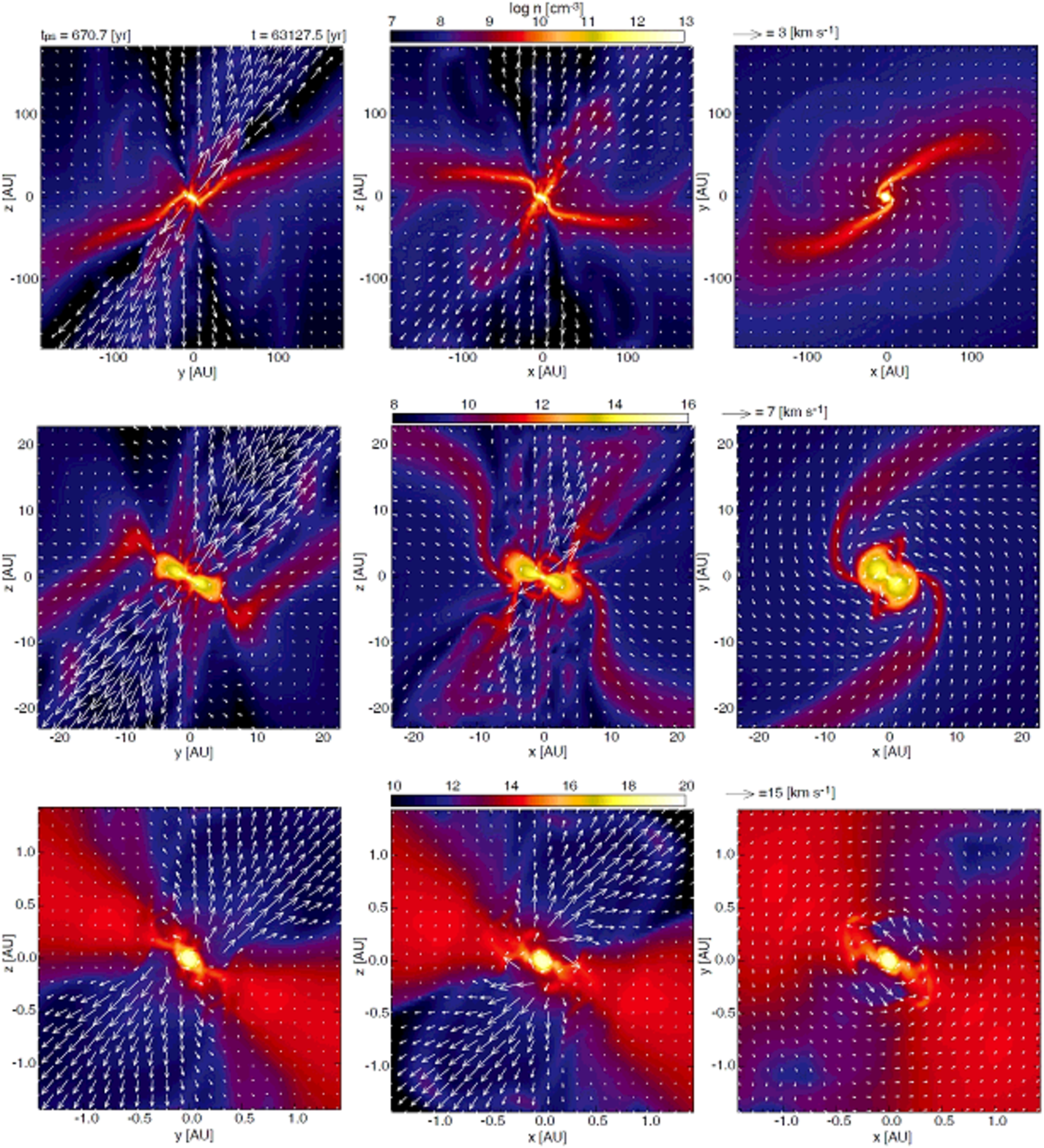}
\end{center}
\caption{
As for Fig.~\ref{fig:1}, but for model T30.
}
\label{fig:7}
\end{figure*}
\clearpage

\begin{figure*}
\begin{center}
\includegraphics[width=1.0\columnwidth]{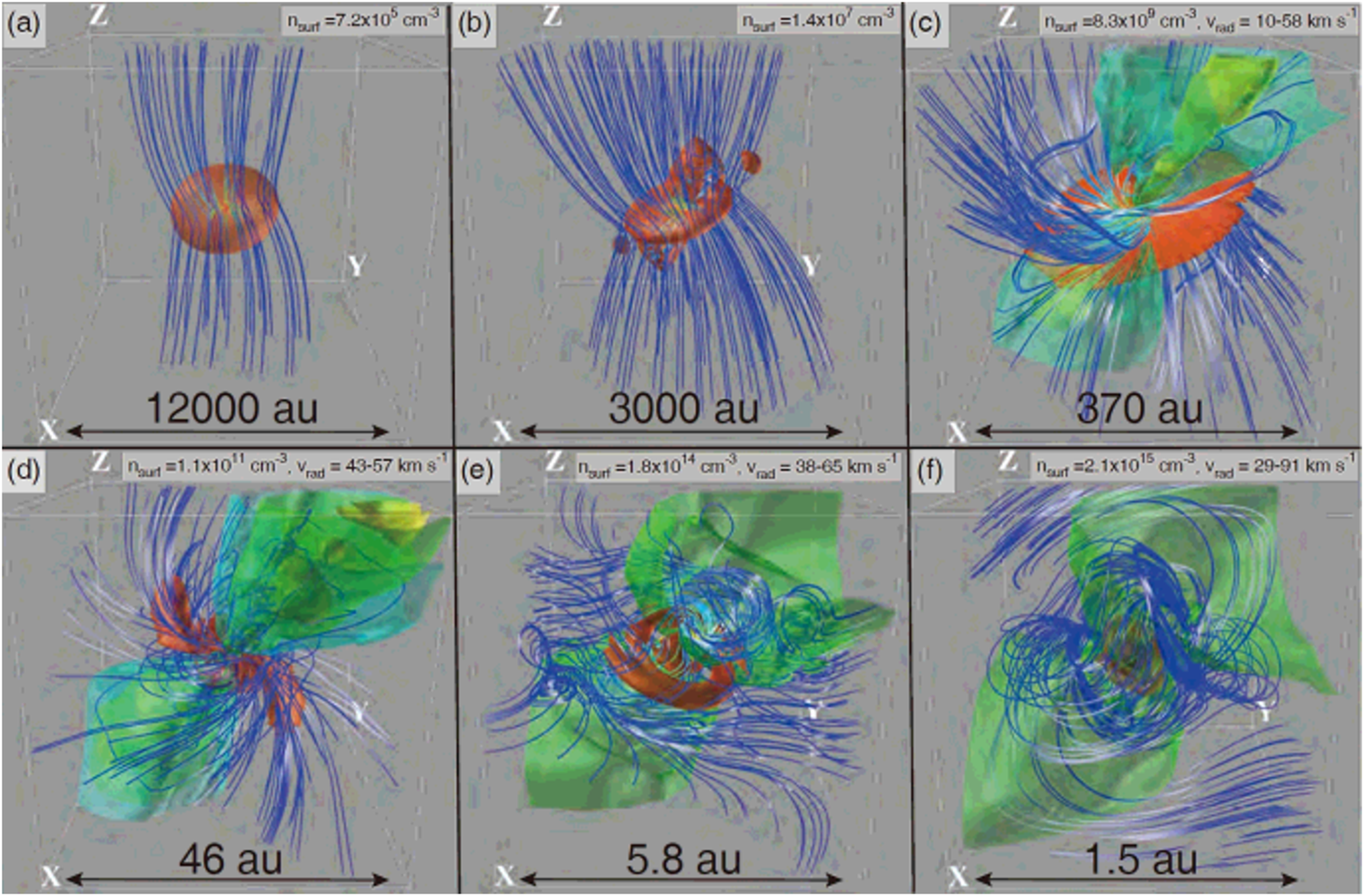}
\end{center}
\caption{
As for Fig.~\ref{fig:2}, but for model T30.
}
\label{fig:8}
\end{figure*}
\clearpage

\clearpage
\begin{figure*}
\begin{center}
\includegraphics[width=0.95\columnwidth]{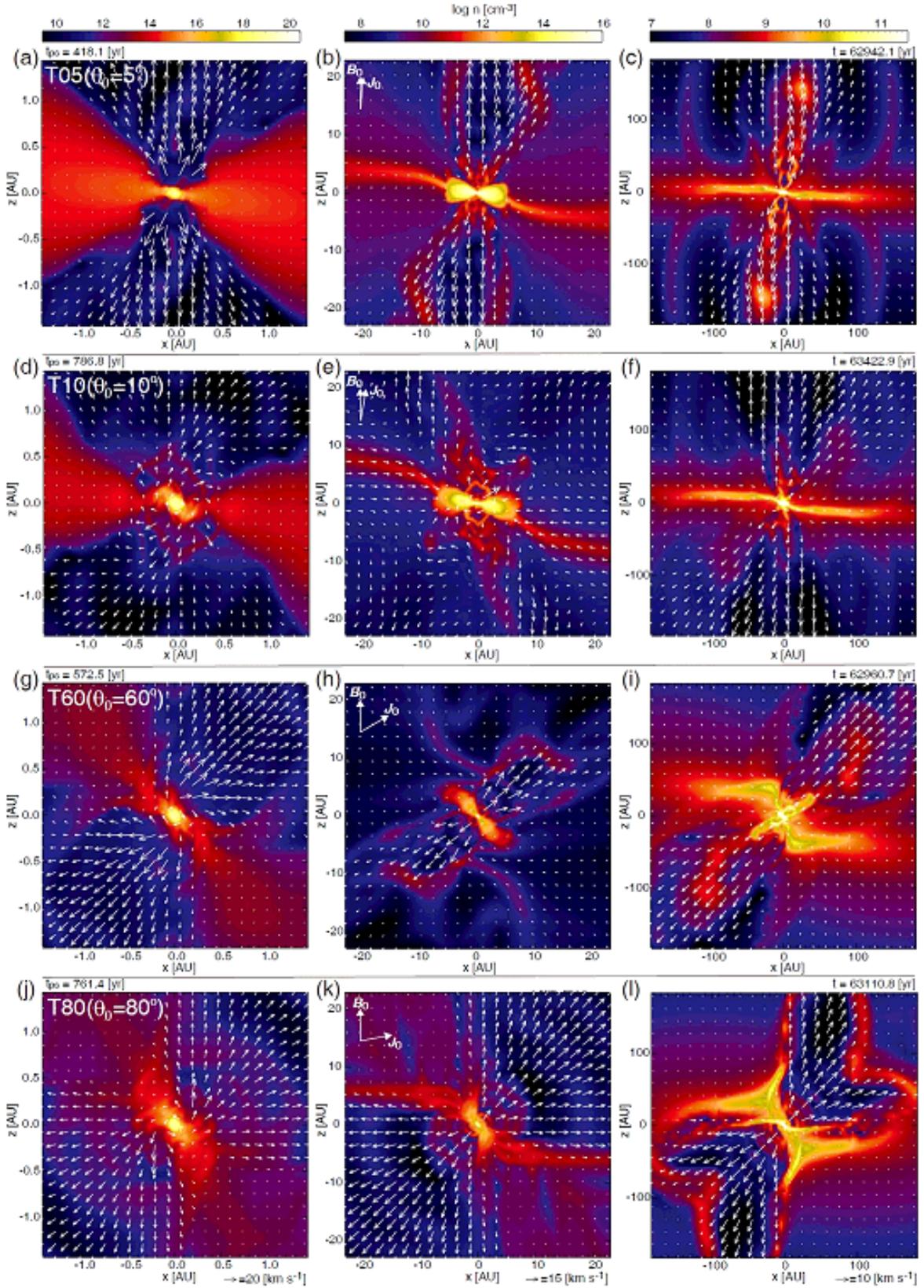}
\end{center}
\caption{
Density (colour) and velocity (arrows) distributions on the $y=0$ plane with a box size  of 2.3\,au (left),  37\,au (middle) and 295\,au (right)  for models T05, T10, T60 and T80.
The elapsed time after protostar formation $t_{\rm ps}$ and that after the beginning of the cloud collapse $t$ are described in the upper part of each panel. 
The model name and parameter $\theta_0$ are noted in each row.
The initial direction of the magnetic field $\vect{B_0}$ and angular momentum $\vect{J_0}$ are represented by arrows in panels ({\it b}), ({\it e}), ({\it h}) and ({\it k}).
}
\label{fig:9}
\end{figure*}
\clearpage

\begin{figure*}
\begin{center}
\includegraphics[width=1.0\columnwidth]{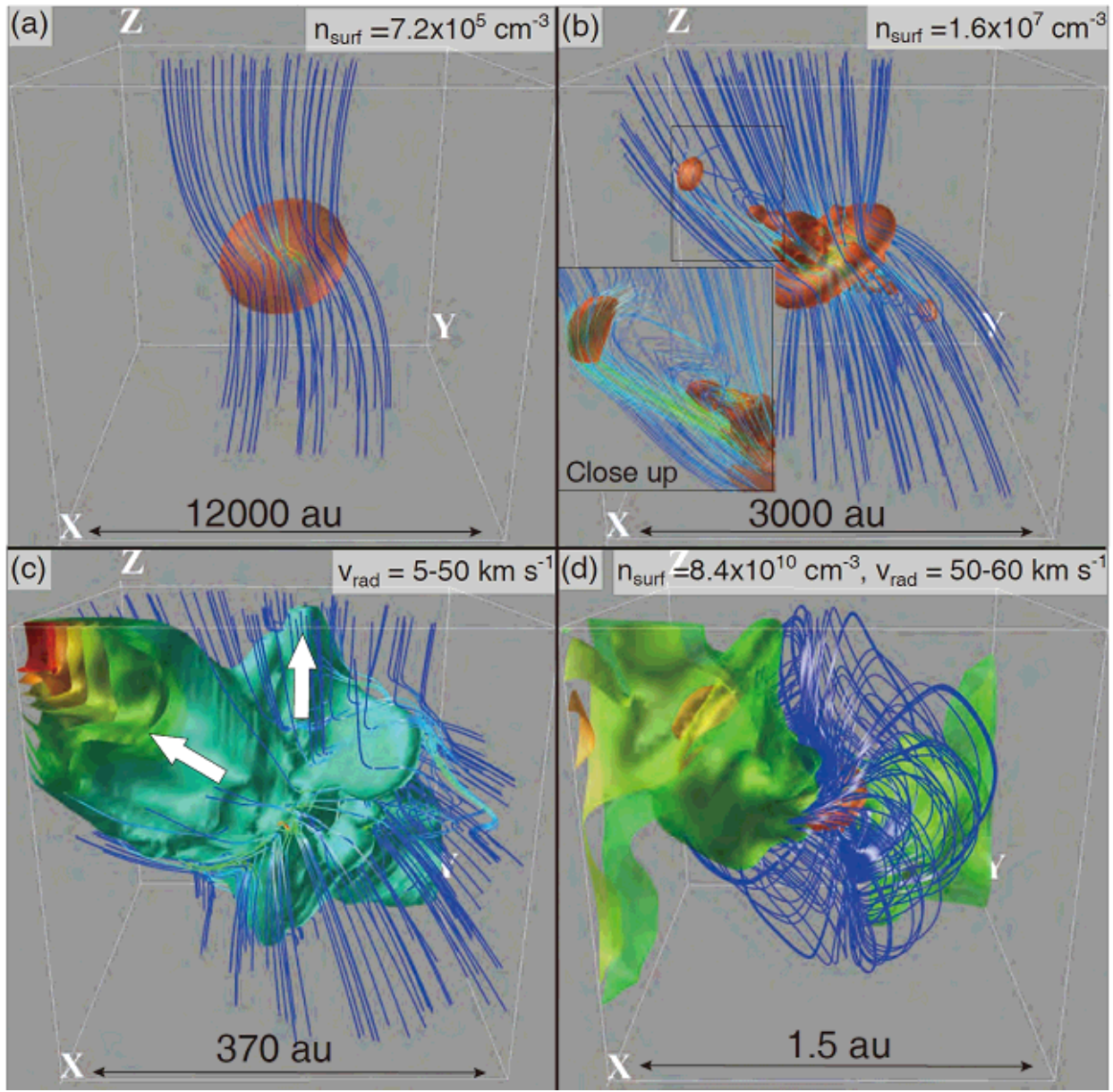}
\end{center}
\caption{
Three-dimensional views of magnetic field lines (blue lines), high-density region (orange iso-density surface around the center) and outflows (blue, green and yellow iso-velocity surfaces) at $t_{\rm ps}=418.1$\,yr and $t=62942.1$\,yr with different spatial scales for model T80.
The box size is denoted in each panel.
The density of the iso-density surface and the velocity range of the outflow  are also described in each panel. 
A close-up view around a knot is inserted in panel ({\it b}).
The outflow directions are denoted by arrows in panel ({\it c}).
}
\label{fig:10}
\end{figure*}
\clearpage

\begin{figure*}
\begin{center}
\includegraphics[width=0.95\columnwidth]{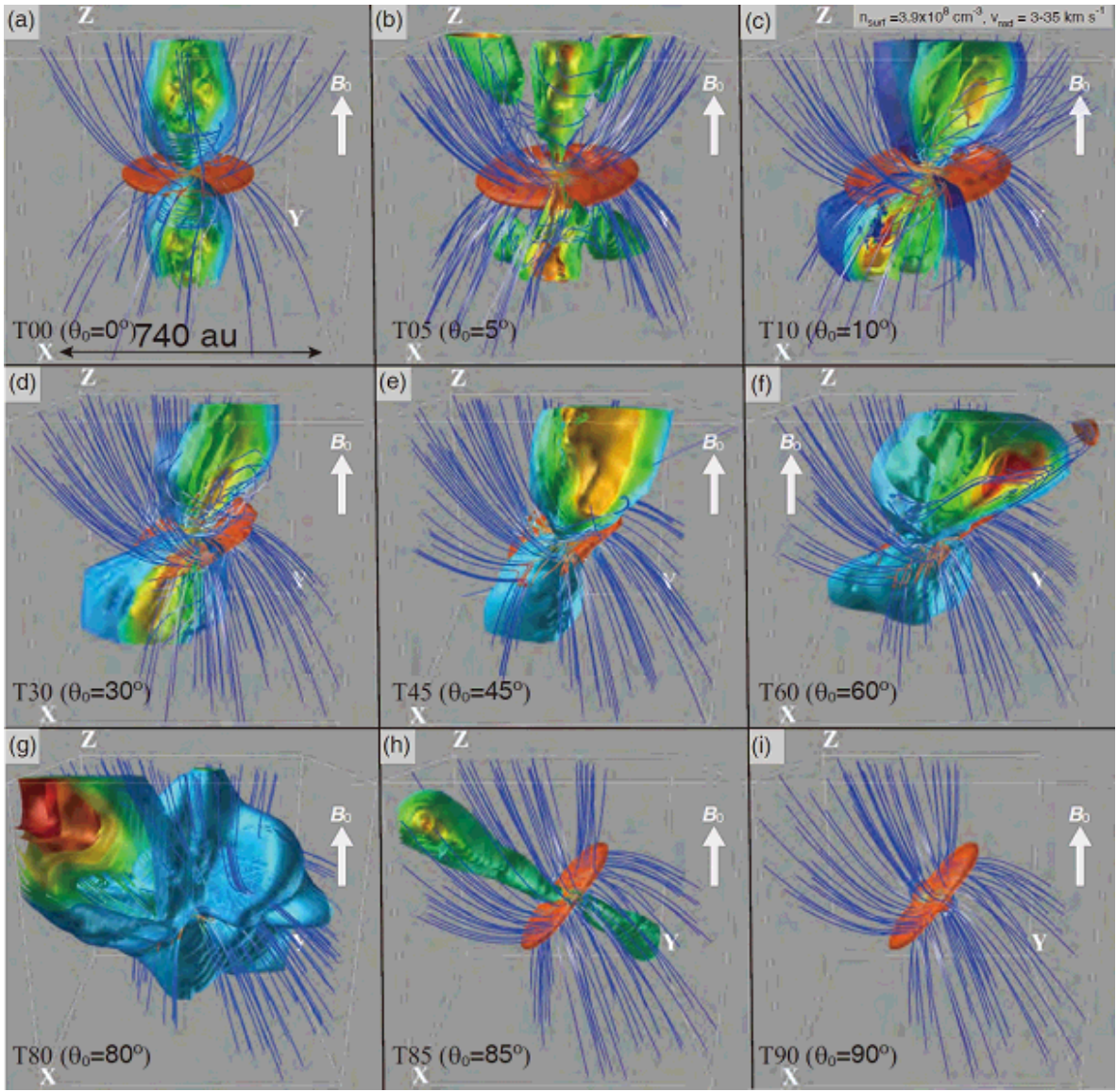}
\end{center}
\caption{
Three-dimensional view of magnetic field lines (blue lines), disc  or high-density region (orange iso-density surface at the center) and outflow (green and yellow surfaces) for all models. 
The box scale is described in panel (a).
The surface density of the iso-density surface $n_{\rm surf}$ and velocity range of the outflow  $v_{\rm rad}$ are described in panel (c).
The box scale, surface density and velocity range are the same in all the panels. 
The initial direction of the magnetic field $\vect{B}_0$ is denoted by an arrow in each panel, in which the initial directions of the magnetic field are also the same among the models.
}
\label{fig:11}
\end{figure*}
\clearpage

\begin{figure*}
\begin{center}
\includegraphics[width=0.95\columnwidth]{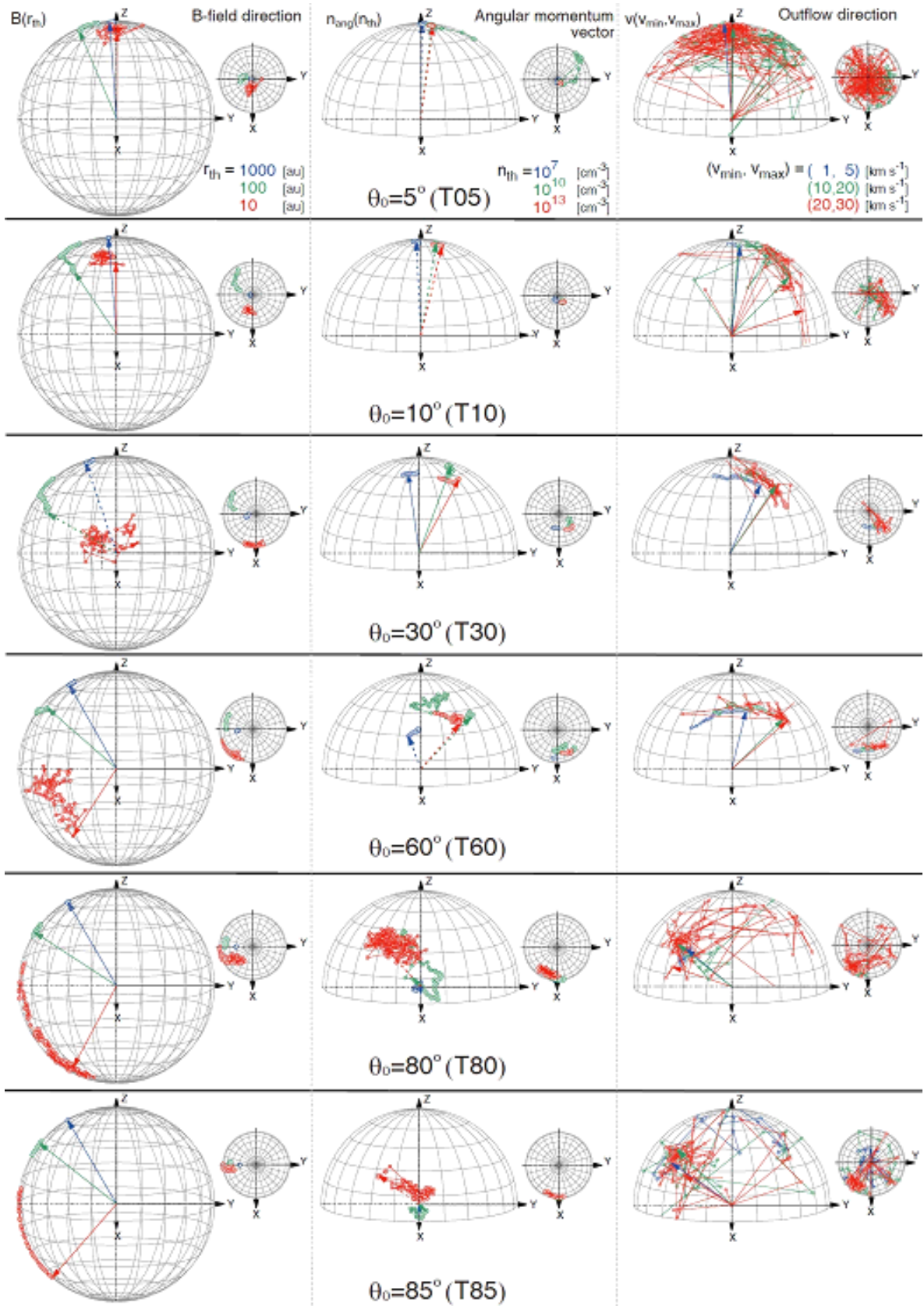}
\end{center}
\caption{
Time evolution of the directions of the magnetic field (left)  angular momentum (middle) and outflow (right) plotted on the surface of a sphere and hemisphere for models T05, T10, T30,  T60, T80 and T85.
The magnetic field directions are measured at three different scales of 1000, 100 and 10\,au.
The angular momentum vectors are measured  for three different threshold densities of $n_{\rm th}=10^7$, $10^{10}$ and $10^{13}\cm$.
The outflows are measured for three different velocity ranges of ($v_{\rm min}$, $v_{\rm max}$)= (1, 5), (10, 20) and (20, 30)$\km$.
The arrows indicate the direction at the end of the simulation.
}
\label{fig:12}
\end{figure*}
\clearpage

\begin{figure*}
\begin{center}
\includegraphics[width=1.0\columnwidth]{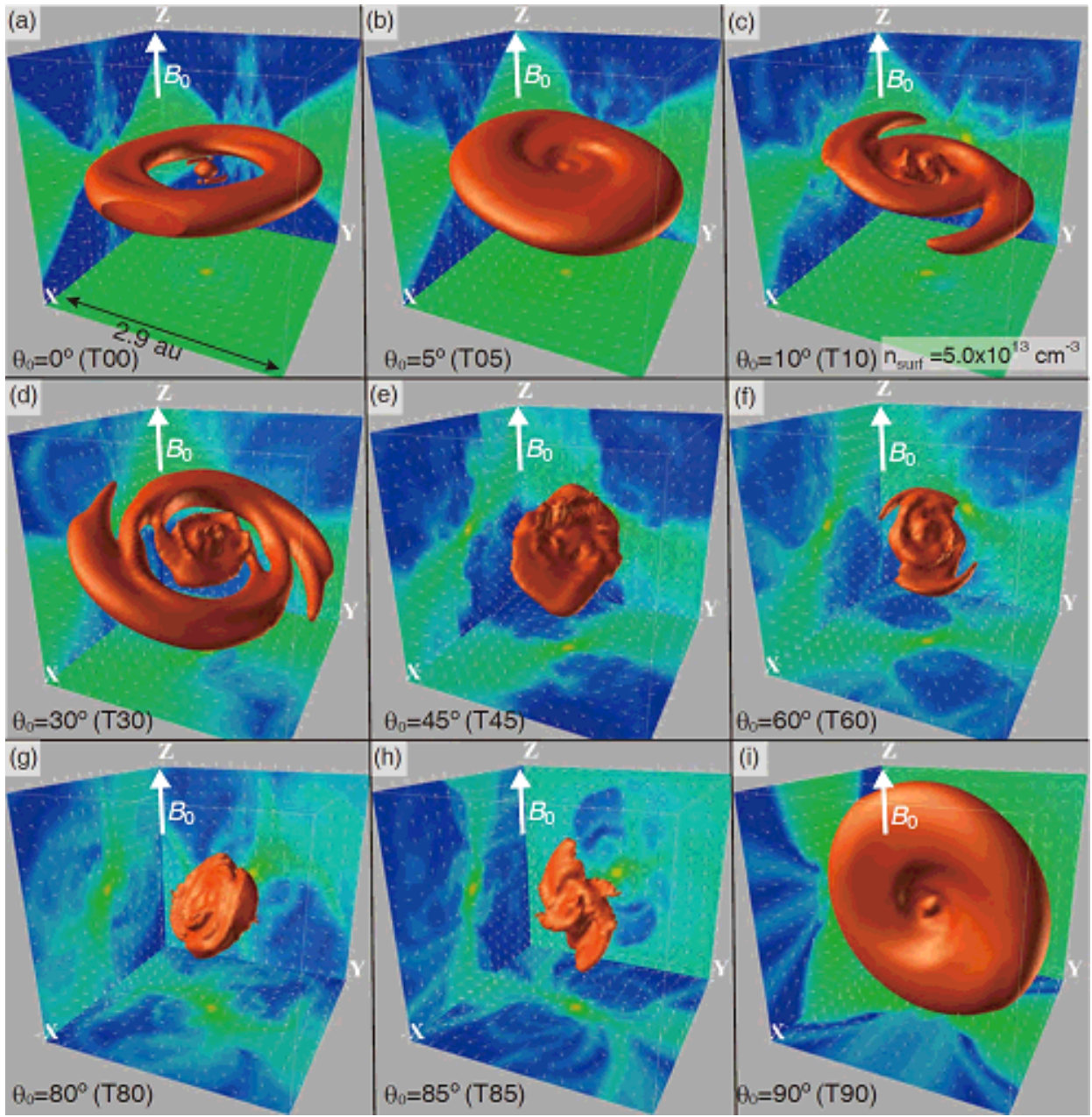}
\end{center}
\caption{
The iso-density surface of $5\times10^{13}\cm$ for all models.
The box scale is  2.9\,au in all the panels.
The density and velocity distributions on the $x=0$, $y=0$ and $z=0$ planes are projected on each wall surface. 
The model name is noted in each panel. 
The initial magnetic vector is indicated by a white arrow in each panel.
}
\label{fig:13}
\end{figure*}
\clearpage

\begin{figure*}
\begin{center}
\includegraphics[width=1.0\columnwidth]{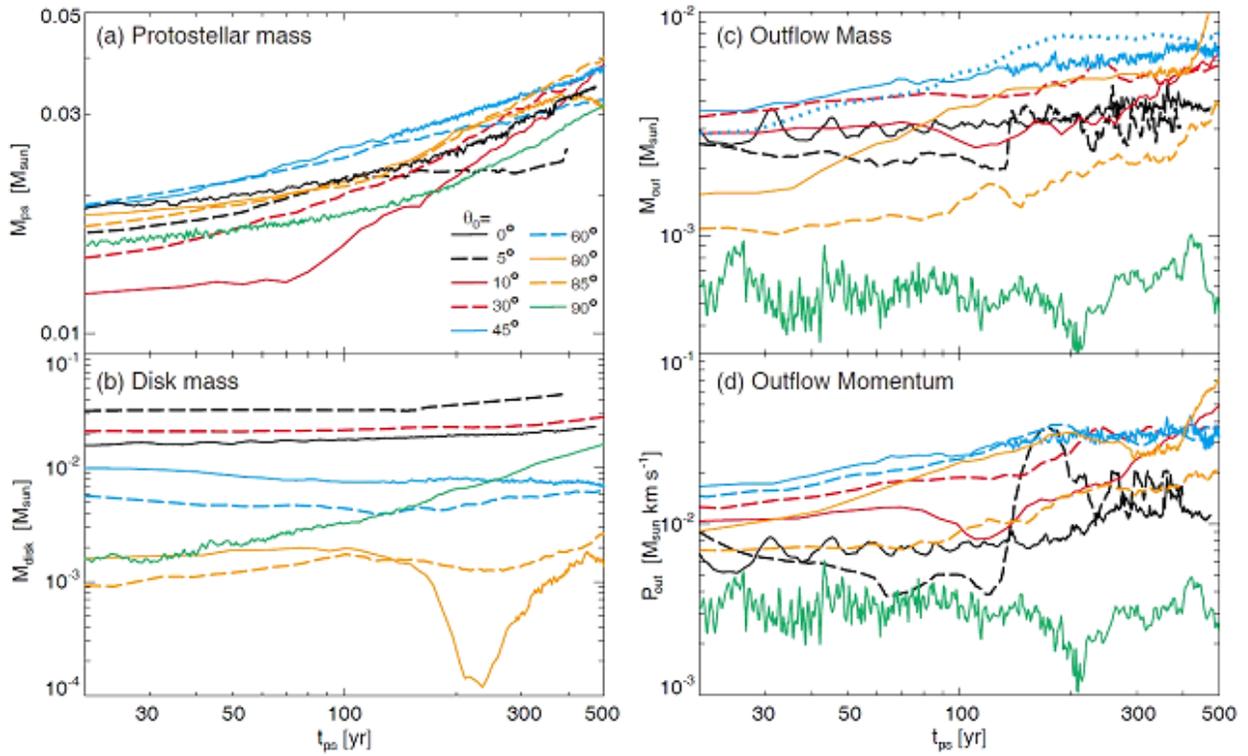}
\end{center}
\caption{
Protostellar mass ({\it a}), disc  mass ({\it b}), outflow mass ({\it c}) and outflow momentum ({\it d}) for all models (T00, T05, T10, T30, T45, T60, T80, T85 and T90) against the elapsed time after protostar formation. 
}
\label{fig:14}
\end{figure*}
\clearpage
\clearpage
\begin{figure*}
\begin{center}
\includegraphics[width=1.0\columnwidth]{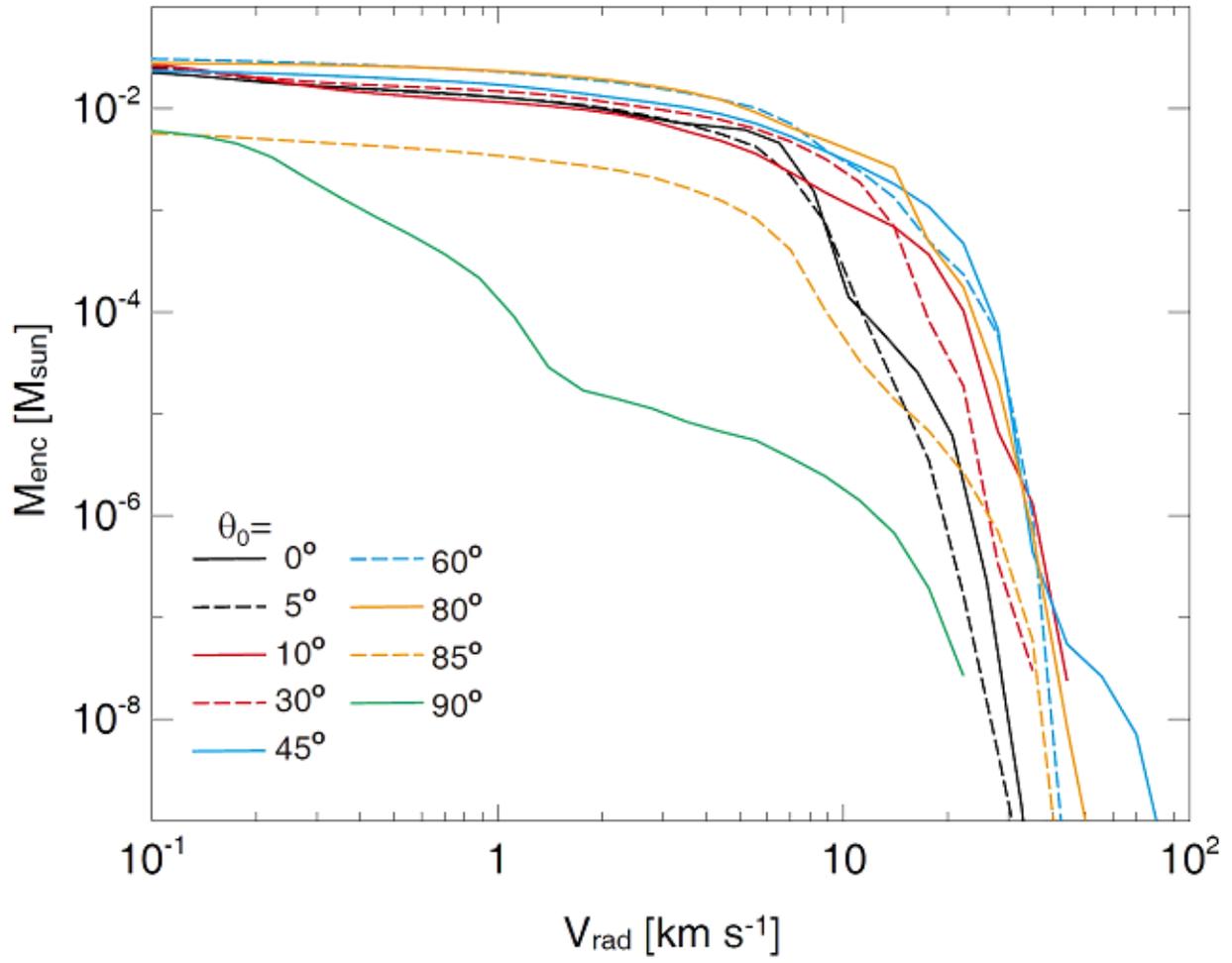}
\end{center}
\caption{
Outflow mass against the outflow velocity at the end of the simulation for all models.
}
\label{fig:15}
\end{figure*}
\end{document}